\documentclass[a4paper,twocolumn]{article} 
\usepackage[top=1.5cm, bottom=1.5cm, left=1.5cm, right=1.5cm, bindingoffset = 0pt, footskip = 0.75cm, marginparwidth = 0pt, marginparsep=0pt]{geometry}
\usepackage{helvet}
\usepackage{upgreek}
\usepackage{graphicx}
\usepackage{float}
\usepackage{caption}
\usepackage[style=nature]{biblatex}
\usepackage{amsmath,amssymb,amsfonts,dsfont}
\usepackage{color}
\usepackage[normalem]{ulem} 
\usepackage{multirow}
\usepackage{array}
\usepackage{mathrsfs}
\usepackage{capt-of}
\usepackage{siunitx}

\newcommand*\diff{\mathop{}\!\mathrm{d}}

\title{Multi-level Quantum Noise Spectroscopy}

\usepackage{authblk}
\author[1,2]{Youngkyu Sung \thanks{youngkyu@mit.edu}}
\author[1]{Antti Veps\"al\"ainen}
\author[1]{Jochen Braum\"uller}
\author[1]{Fei Yan}
\author[1]{Joel I-Jan Wang}
\author[1]{Morten Kjaergaard}
\author[1]{Roni Winik}
\author[1]{Philip Krantz}
\author[1]{Andreas Bengtsson}
\author[3]{Alexander J. Melville}
\author[3]{Bethany M. Niedzielski}
\author[3]{Mollie E. Schwartz}
\author[3]{David K. Kim}
\author[3]{Jonilyn L. Yoder}
\author[1,2]{Terry P. Orlando}
\author[1]{Simon Gustavsson}
\author[1,2,3,4]{\mbox{William D. Oliver} \thanks{william.oliver@mit.edu}}

\affil[1]{Research Laboratory of Electronics, Massachusetts Institute of Technology, Cambridge, MA 02139, USA}
\affil[2]{Department of Electrical Engineering and Computer Science, Massachusetts Institute of Technology, Cambridge, MA 02139, USA}
\affil[3]{MIT Lincoln Laboratory, 244 Wood Street, Lexington, MA 02421, USA}
\affil[4]{Department of Physics, Massachusetts Institute of Technology, Cambridge, MA 02139, USA}

\date{\today}

\begin{document}
\maketitle
\textsf{\bfseries{ 
System noise identification is crucial to the engineering of robust quantum systems. Although existing quantum noise spectroscopy (QNS) protocols measure an aggregate amount of noise affecting a quantum system, they generally cannot distinguish between the underlying processes that contribute to it. Here, we propose and experimentally validate a spin-locking-based QNS protocol that exploits the multi-level energy structure of a superconducting qubit to achieve two notable advances. First, our protocol extends the spectral range of weakly anharmonic qubit spectrometers beyond the present limitations set by their lack of strong anharmonicity. Second, the additional information gained from probing the higher-excited levels enables us to identify and distinguish contributions from different underlying noise mechanisms. 
}}
\\

\section*{Introduction}
Studying noise sources affecting quantum mechanical systems is of great importance to quantum information processing, quantum sensing applications, and the fundamental understanding of microscopic noise mechanisms~\cite{Preskill2018, Degen2017, Paladino2014, Krantz2019}. 
Generally, a quantum two-level system -- a qubit -- is employed as a sensor of noise that arises from the qubit environment including both classical and quantum sources~\cite{Degen2017,Schoelkopf2002}. 
By driving the qubit with suitably designed external control fields and measuring its response in the presence of environmental noise, the spectral content of the noise can be extracted~{\cite{Alvarez2011, Yuge2011, Young2012, Paz-Silva2014, Ithier2005}}. Such noise spectroscopy techniques are generally s ferred to as quantum noise spectroscopy (QNS) protocols.
Over the past two decades, QNS protocols have been explored for both pulsed (free-evolution) and continuous (driven-evolution) control schemes and experimentally implemented across many qubit platforms -- including  diamond nitrogen vacancy centers~\cite{Meriles2010,Romach2015}, {nuclear spins~\cite{Vandersypen05,Alvarez2011}, cold atoms~\cite{ Carter2013}, superconducting quantum circuits~\cite{Ithier2005,Bylander2011,Yan2013,Yoshihara2014,Quintana2017}}, semiconductor quantum dots~\cite{Dial2013,Muhonen2014,Morello2018,Tarucha2018}, and trapped ions \cite{Frey2017}. 
Although these protocols have generally focused on Gaussian noise models, a new QNS protocol was recently developed and demonstrated that enables higher-order spectral estimation of non-Gaussian noise in quantum systems~\cite{Norris2016,Sung2019}.

Since QNS protocols commonly presume a qubit platform, they have generally been developed within a two-level system approximation, without regard for higher energy levels.
As a consequence, despite tremendous progress and successes, QNS protocols have certain limitations (for example, limited bandwidth) when applied to weakly anharmonic qubits such as the transmon~\cite{Koch2007, Majer2007}, the gatemon~\cite{Larsen2015, Wang2019}, or the capacitively shunted flux qubit~\cite{Yan2016}. 
However, since weakly anharmonic superconducting qubits are among the most promising platforms being considered for realizing quantum information processors~\cite{Arute2019}, noise spectroscopy techniques that incorporate the effects of higher-excited states in these qubits must be developed to further improve their coherence and gate performance.

Among existing QNS protocols, the spin-locking approach has been shown to be applicable to both classical and non-classical noise spectra. 
It is also experimentally advantageous, using a relatively straightforward relaxometry analysis to extract a spectral decomposition of the environmental noise affecting single qubits~\cite{Yan2013,Yan2018}, and it has recently been extended to measure the cross-spectra of spatially correlated noise in multi-qubit systems~\cite{Luepke2020}.
{As with many contemporary QNS protocols, the spin-locking approach presumes a two-level-system approximation. While this approximation holds at low frequencies, its validity breaks down as one attempts to perform noise spectroscopy at frequencies approaching and exceeding qubit anharmonicity (e.g., around 200 -- 300 MHz is conventional superconducting transmon qubits) due to the impact of additional energy levels, leading to systematic errors in the extracted noise spectrum.}

{In this work, we develop a multi-level spin-locking QNS protocol  and experimentally validate it using a flux-tunable transmon qubit and accounting for five energy levels.} 
We demonstrate an accurate spectral reconstruction of engineered flux noise over a frequency range $\SI{50}{\mega\hertz}$ to $\SI{300}{\mega\hertz}$, overcoming the spectral limitations imposed by the sensor's relatively weak anharmonicity of approximately $\SI{200}{\mega\hertz}$. 
Furthermore, by measuring the power spectra of dephasing noise acting on both the $|0\rangle$--$|1\rangle$ and $|1\rangle$--$|2\rangle$ transitions, we extract and uniquely identify noise contributions from both flux noise and photon shot noise, an attribution that is not possible within solely a two-level approximation.

{
\section*{Results}
}

{
\subsection*{System Description}
}
We consider an externally-driven $d$-level quantum system $(d>2)$, which serves as the quantum noise sensor that evolves under the influence of its noisy environment (bath). 
Throughout this work, we consider only pure dephasing ($\sigma_z$-type) noise.
The impact of energy relaxation ($T_1$) on our protocol is discussed in {Supplementary Note}~6. In the interaction picture with respect to the bath Hamiltonian $H_{\mathrm{B}}$, the joint sensor-environment system 
can be described by the Hamiltonian:
\begin{align}
    H(t) =& \hbar \sum_{j=1}^{d-1} \Big [ \left(\omega_{\mathrm{s}}^{(j)} +B^{(j)}(t) \right) |j \rangle \langle j| 
    \nonumber \\
    &+\xi(t) \lambda^{(j-1,j)} (\sigma_{+}^{(j-1,j)} + \sigma_{-}^{(j-1,j)}) \Big ],
    \label{eq:H_lab}
\end{align}
where $|j \rangle \langle j|$ is the projector for the $j$-th level of the multi-level sensor. 
The sensor eigenenergies are $\hbar \omega_{\mathrm{s}}^{(j)}$ with the ground state energy set to zero, and the ${B}^{(j)}(t)$ correspond to the time-dependent noise operators that longitudinally couple to the $j$-th level of the sensor and cause level $j$ to fluctuate in energy.
The raising and lowering operators of the sensor are denoted by
$\sigma_{+}^{(j-1,j)}\equiv|j\rangle \langle j-1|$ and  $\sigma_{-}^{(j-1,j)}\equiv|j-1\rangle \langle j|$, respectively. 
The external driving field is denoted by $\xi(t)$. 
We continuously drive the multi-level sensor with a signal 
\begin{align}
\xi(t) = A_{\mathrm{drive}} \cos(\omega_{\mathrm{drive}}t + \phi),
\end{align}
where $A_{\mathrm{drive}}$, $\omega_{\mathrm{drive}}$, and $\phi$ correspond to the amplitude, the frequency and the phase of the driving field, respectively, and we assume $\phi = 0$ without loss of generality. 
The parameter $\lambda^{(j-1,j)}$ represents the strength of the $|j-1\rangle $--$|j\rangle$ transition relative to the $|0\rangle$--$|1\rangle$ transition with  $\lambda^{(0,1)} \equiv 1$.

When the drive frequency $\omega_{\mathrm{drive}}$ is resonant with the $|0\rangle$--$|1\rangle$ transition frequency $\omega_{\mathrm{s}}^{(0,1)}$ of the sensor, the first two levels form a pair of dressed states, $|+^{(0,1)}\rangle$ and $|-^{(0,1)}\rangle$. 
The level separation between dressed states is the Rabi frequency $\Omega^{(0,1)}$, and it is determined predominantly (although not exactly, as we describe below) by the effective driving strength $\lambda^{(0,1)} A_{\mathrm{drive}} \equiv A_{\mathrm{drive}}$~\cite{Bishop2009,Braumuller15}. 
These dressed states form the usual spin-locking basis $\{ |+^{(0,1)}\rangle, |-^{(0,1)}\rangle \}$ of a conventional, driven two-level sensor~\cite{Yan2013,Yan2018,Luepke2020}. 

\begin{figure}[t!]
    \centering
    \includegraphics[width=8.6cm]{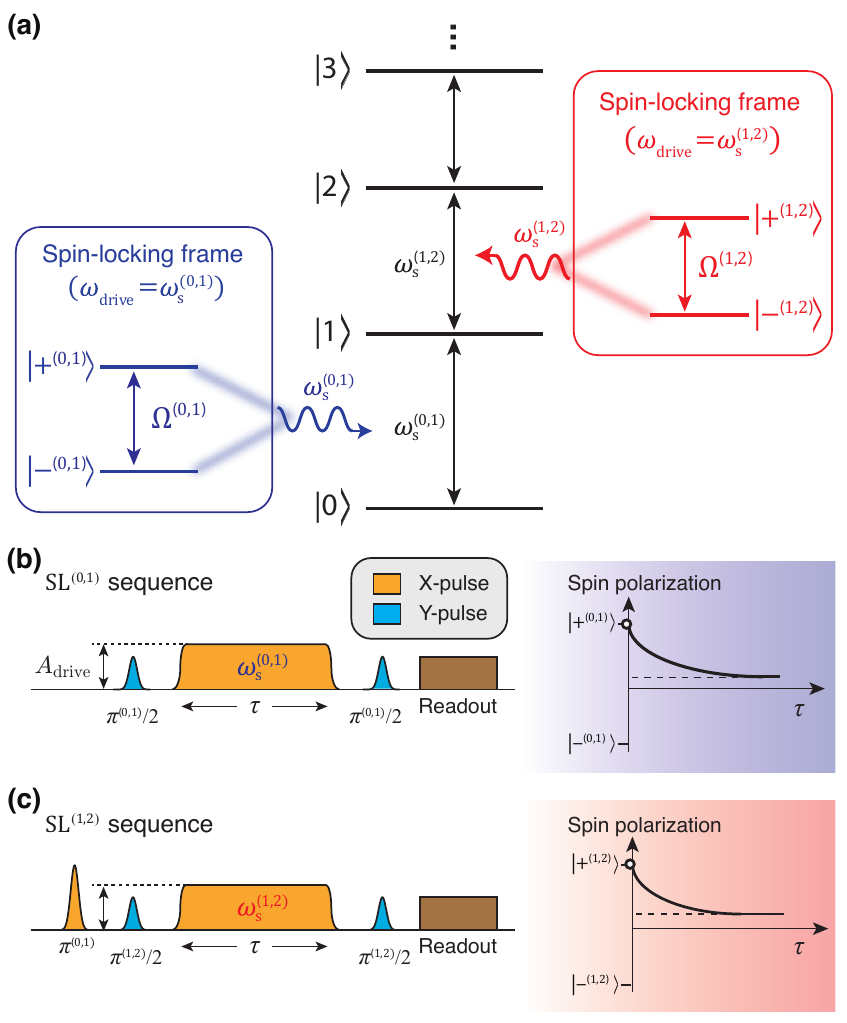}
    \caption{Spin-locking noise spectroscopy in a multi-level sensor. 
    \textbf{(a)} A transition between the $(j-1)$-th and $j$-th level of a multi-level system is driven resonantly to form the $j$-th spin-locking basis (dressed states) \{$|+^{(j-1,j)}\rangle$,  $|-^{(j-1,j)}\rangle$\} which are separated by the Rabi frequency $\Omega^{(j-1,j)}$. The two-level system formed by the basis \{$|+^{(j-1,j)}\rangle$,  $|-^{(j-1,j)}\rangle$\} acts as the $j$-th spin-locked noise spectrometer.
    \textbf{(b)-(c)} Spin-locking (SL) sequences used to measure the relaxation of spin polarization $|+^{(0,1)}\rangle$ and $|+^{(1,2)}\rangle$ as a function of the spin-locking duration $\tau$, respectively. 
    }
    \label{fig:Fig1}
\end{figure}

We now generalize the two-level spin-locking concept to the case of a multi-level sensor. 
By resonantly driving at the frequency $\omega_{\mathrm{s}}^{(j-1,j)}\equiv \omega_{\mathrm{s}}^{(j)}-\omega_{\mathrm{s}}^{(j-1)}$ of the transition between states $|j-1\rangle$ and $|j\rangle$, the system forms a pair of dressed states, $|+^{(j-1,j)}\rangle$ and $|-^{(j-1,j)}\rangle$, separated by a Rabi frequency $\Omega^{(j-1,j)}$~(see Fig.~\ref{fig:Fig1}.(a)) that is determined predominantly by an effective driving strength $\lambda^{(j-1,j)} A_{\mathrm{drive}}$. 
The effective two-level system formed by the basis $\{ |+^{(j-1,j)}\rangle,|-^{(j-1,j)}\rangle\}$ acts as the $j$-th spectrometer and probes dephasing noise that leads to a fluctuation of the $|j-1\rangle$--$|j\rangle$ transition at frequency $\Omega^{(j-1,j)}$. The case $j=1$ then corresponds back to the conventional two-level noise sensor.

Throughout the main text, we will refer to the reference frame and two-dimensional subspace defined by the $j$-th spin-locking basis $\{ |+^{(j-1,j)}\rangle$, $|-^{(j-1,j)}\rangle \}$ as the $j$-th \textit{spin-locking frame} and the $j$-th \textit{spin-locking subspace}, respectively. 
To move to the $j$-th spin-locking frame, we apply unitary transformations and truncate the Hilbert space of the multi-level sensor into the $j$-th spin-locking subspace (see detailed derivation in {Supplementary Note}~4). Then, the effective Hamiltonian describing the $j$-th noise spectrometer is: 
\begin{align}
    \label{eq:H_SL_eff}
    \tilde{H}_{\mathrm{SL}}^{(j-1,j)}(t) =& \frac{\hbar}{2} \left[ \Omega^{(j-1,j)} + \tilde{B}^{(j-1,j)}_{\parallel}(t) \right] \tilde{\sigma}_z^{(j-1,j)}
    \nonumber \\ 
    &+\hbar  \tilde{B}^{(j-1,j)}_{\perp} (t) \left(\tilde{\sigma}_{+}^{(j-1,j)} +\tilde{\sigma}_{-}^{(j-1,j)}\right), 
\end{align}
where $\tilde{\sigma}_z^{(j-1,j)}$, $\tilde{\sigma}_+^{(j-1,j)}$, and $\tilde{\sigma}_-^{(j-1,j)}$ denote the Pauli Z operator, the raising operator, and the lowering operator of the $j$-th spin-locked spectrometer, respectively.  
{
The longitudinal noise in the lab frame (Eq.~(\ref{eq:H_lab})) for a multi-level system leads to both transverse and longitudinal noise in the spin-locking frame. As a result, the longitudinal noise operator $B^{(j)}(t)$ in the lab frame is transformed into the spin-locking frame as a transverse noise operator $\tilde{B}^{(j-1,j)}_{\perp}(t)$, which leads to longitudinal relaxation, and the longitudinal noise operator $\tilde{B}^{(j-1,j)}_{\parallel}(t)$, which leads to transverse relaxation, within the $j$-th spin locking subspace.}
They are given as linear combinations of $B^{(j)}(t)$, arising from the level dressing across multiple levels as follows:
\begin{align}
    \label{eq:B_perp_general}
    \tilde{B}^{(j-1,j)}_{\perp}(t) =& \sum_{k=1}^{d-1} \alpha^{(k)}_{(j-1,j)} B^{(k)}(t), 
    \\ 
    \label{eq:B_parallel_general}
    \tilde{B}^{(j-1,j)}_{\parallel}(t) =& \sum_{k=1}^{d-1} \beta^{(k)}_{(j-1,j)} B^{(k)}(t),
\end{align}
where we define the \emph{noise participation ratio} $\alpha^{(k)}_{(j-1,j)}$ ($\beta_{(j-1,j)}^{(k)}$) as a dimensionless factor that quantifies the fraction of the dephasing noise at the $k$-th level that is transduced (i.e., projected) to transverse (longitudinal) noise of the $j$-th pair of spin-locked states. 
The noise participation ratios $\alpha^{(k)}_{(j-1,j)}$ and $\beta_{(j-1,j)}^{(k)}$ can be estimated by numerically solving for the dressed states in terms of the bare states $|j\rangle$ (see {Supplementary Note}~4 for details). 
Note that the sign of the noise participation ratios can be either positive or negative, leading to the possibility for effective constructive and destructive interference between the noise operators $B^{(k)}(t)$.

There are two noteworthy distinctions between a manifestly two-level system and a multi-level system. 
First, although the splitting energy $\hbar\Omega^{(j-1,j)}$ between the $j$-th pair of dressed states ({the $j$-th} spin-locked states) is \emph{predominantly} determined by the effective driving energy $\hbar(\lambda^{(j-1,j)} A_{\mathrm{drive}})$, they are not universally equivalent.  
For an ideal two-level system within the rotating wave approximation, the Rabi frequency is indeed proportional to the effective driving field via the standard Rabi formula~\cite{Yan2013,Yan2018,Luepke2020}. 
However, this is not generally the case in a multi-level setting due to additional level repulsion from adjacent dressed states~\cite{Koshino2013}. Rather, in the multi-level setting of relevance here, the distinction between $\Omega^{(j-1,j)}$ and $\lambda^{(j-1,j)}A_{\mathrm{drive}}$ must be taken into account to yield an accurate estimation of the noise spectrum. 

Second, as a consequence of the multi-level dressing, more than two noise operators $B^{(k)}(t)$ generally contribute to the longitudinal relaxation within a given pair of spin-locked states. In the limit where $\lambda^{(j-1,j)}A_{\mathrm{drive}}$ is small compared to the sensor anharmonicities, the Eqs.~(\ref{eq:B_perp_general})-(\ref{eq:B_parallel_general}) reduce to 
\begin{align}
\label{eq:B_small_Omega}
    \tilde{B}^{(j-1,j)}_{\perp}(t) \approx \frac{1}{2} \left[B^{(j-1)}(t) - B^{(j)}(t) \right],
    \hspace{3mm}
    \tilde{B}^{(j-1,j)}_{\parallel}(t) \approx 0, 
\end{align}
which conform to the standard spin-locking noise spectroscopy protocol for a two-level sensor~\cite{Yan2013,Yan2018,Luepke2020}. 
However, as the effective drive strength $\lambda^{(j-1,j)}A_{\mathrm{drive}}$ increases, the contribution of peripheral bare states -- \emph{other} than $|j-1\rangle$ and $|j\rangle$ -- to the formation of the spin-locked states $|+^{(j-1,j)}\rangle$ and $|-^{(j-1,j)}\rangle$ increases. 
As a result, in the large $\lambda^{(j-1,j)} {A_{\mathrm{drive}}}$ limit, the multi-level dressing transduces the frequency fluctuations of more than two levels to the longitudinal relaxation within the $j$-th spin locking frame. Also, this multi{-}level effect contributes to the emergence of non-zero transverse relaxation $B_{\parallel}^{(j-1,j)}(t)$, terms which would otherwise be absent within a two-level approximation~\cite{Yan2013, Yan2018,Luepke2020}. \\

{
\subsection*{Noise Spectroscopy Protocol}
}
The multi-level noise spectroscopy protocol introduced here consists of 
measuring the energy decay rate $\Gamma^{(j-1,j)}_{1\rho}$ (\emph{i.e.,} longitudinal relaxation rate) 
and the polarization $\langle \tilde{\sigma}_z^{(j-1,j)}(\tau) \rangle$ 
in the $j$-th spin-locking frame, and then uses these quantities to extract the spectral density $\tilde{S}_{\perp}^{(j-1,j)}$ of the noise transverse to the spin-locking quantization axis. 
This in turn can be related to the longitudinal spectral density $S_{\parallel}^{(j-1,j)}$ that causes dephasing (\emph{i.e.,} transverse relaxation) in the original, undriven reference frame (the qubit ``lab frame'' \cite{Yan2013}). 

We begin by preparing the multi-level sensor in the $j$-th spin-locked state $|+^{(j-1,j)}\rangle$ by applying a sequence of resonant $\pi$ pulses $\left[\pi^{(0,1)}, \pi^{(1,2)}, \cdots \pi^{(j-2,j-1)}\right]$, which act to sequentially excite the sensor from the ground state $|0\rangle$ to state $|j-1\rangle$. 
We then apply a $\pi^{(j-1,j)}/2$ pulse along the $y$-axis of the Bloch sphere, where the north and south poles now correspond to $|j-1\rangle$ and $|j\rangle$, respectively~{\cite{Vandersypen05}}. The pulse acts to rotate the Bloch vector from the south pole to the $x$-axis, thereby placing 
the multi-level sensor in the $j$-th spin-locked state $|+^{(j-1,j)}\rangle = (|j-1\rangle + |j\rangle)/\sqrt{2}$. 
Subsequently, a spin-locking drive with amplitude $A_{\mathrm{drive}}$ is applied along the $x$-axis (collinear with the Bloch vector) at a frequency resonant with the  $|j-1\rangle$-$|j\rangle$ transition and for a duration $\tau$. 
By adiabatically turning on and off the drive, we keep the state of the sensor within the $j$-th spin-locking subspace. 
Once the drive is off, a second $\pi^{(j-1,j)}/2$ pulse is applied along the $y$-axis in order to map the spin-locking basis $\{|+^{(j-1,j)}\rangle, |-^{(j-1,j)}\rangle \}$ onto the measurement basis $\{|j\rangle, |j-1\rangle\}$, and the qubit is then read out.
This procedure is then repeated $N$ times to obtain estimates for the probability of being in states $\{|j\rangle$ and $|j-1\rangle\}$, which represent the probability of being in states $\{|+^{(j-1,j)}\rangle$ and $|-^{(j-1,j)}\rangle \}$, respectively.

The above protocol is then repeated as a function of $\tau$ in order to measure the longitudinal spin-relaxation decay-function of the $j$-th spin-locked spectrometer. For each $\tau$, we define a normalized polarization of the spectrometer, 
\begin{align}
    \langle \tilde{\sigma}_z^{(j-1,j)}(\tau) \rangle \equiv \frac{\rho^{(j-1,j-1)}(\tau) - \rho^{(j,j)}(\tau) }{\rho^{(j-1,j-1)}(\tau) + \rho^{(j,j)}(\tau)},
\end{align}
where $\rho^{(j,j)}(\tau)$ denotes the population (the probability) of the $j$-th level.
From the $\tau$-dependence of $\langle \tilde{\sigma}_z^{(j-1,j)}(\tau) \rangle$, we extract both the relaxation rate $\Gamma^{(j-1,j)}_{1\rho}$ of the spin polarization and the equilibrium polarization $\tilde{\sigma}_z^{(j-1,j)}(\tau)|_{\tau\rightarrow\infty}$. 
{The values $\Gamma^{(j-1,j)}_{1\rho}$ and $\tilde{\sigma}_z^{(j-1,j)}(\tau)|_{\tau\rightarrow\infty}$ extracted from an experiment performed at a particular Rabi frequency $\Omega^{(j-1,j)}$ are related to the transverse noise PSD $\tilde{S}_{\perp}^{(j-1,j)}(\omega)$ at angular frequency $\omega = \Omega^{(j-1,j)}$} as follows~({see Supplementary Note}~5 for details):
\begin{align}
    \label{eq:gamma_1rho}
    \Gamma_{1\rho}^{(j-1,j)} = \tilde{S}_{\perp}^{(j-1,j)} ({\omega}) +\tilde{S}_{\perp}^{(j-1,j)} (-{\omega}), 
\end{align}
\begin{align}
    \label{eq:equilibrium_state_polarization}
    \langle \tilde{\sigma}_z^{(j-1,j)}(t) \rangle |_{t\rightarrow\infty} = \frac{\tilde{S}_{\perp}^{(j-1,j)} ({\omega}) -\tilde{S}_{\perp}^{(j-1,j)} (-{\omega})}{\tilde{S}_{\perp}^{(j-1,j)} ({\omega}) +\tilde{S}_{\perp}^{(j-1,j)} (-{\omega})}.
\end{align}
Here, the transverse noise spectrum $\tilde{S}_{\perp}^{(j-1,j)}({\omega})$ 
is the Fourier transform of the two-time correlation function of the transverse noise operators acting on the spectrometer:
\begin{align}
    \tilde{S}_{\perp}^{(j-1,j)}({\omega}) &= \int_{-\infty}^{\infty}\diff{\tau} e^{-i{\omega} \tau} \langle \tilde{B}^{(j-1,j)}_{\perp}(\tau)\tilde{B}^{(j-1,j)}_{\perp}(0)\rangle.
\end{align}

In the following noise spectroscopy measurements, we will record the spin relaxation for the 1'st and 2'nd spin-locked noise spectrometers [Fig.~\ref{fig:Fig1}(b) and (c)]. 
Then, the traces are fit to an exponential decay, allowing us to extract $\Gamma_{1\rho}^{(j-1,j)}$ and $\langle \tilde{\sigma}_z^{(j-1,j)}(t)\rangle |_{t\rightarrow\infty}$.
This is repeated for various drive amplitude $A_{\mathrm{drive}}$ in order to reconstruct $\tilde{S}_{\perp}^{(j-1,j)} ({\omega})$. %
For simplicity, we will hereafter refer to the  spin-locking noise spectroscopy exploiting the $|j-1\rangle$--$|j\rangle$ transition as $\mathrm{SL}^{(j-1,j)}$.
To validate the protocol, we will perform the spin relaxation experiments both in the presence and in the absence of engineered noise, and 
distinguish the contributions of $T_1$ decay and native dephasing noise from the estimation of $\tilde{S}^{(j-1,j)}_{\perp}({\omega})$ {(see Supplementary Note~7 for details)}. \\

\begin{figure}[t!]
    \centering
    \includegraphics[width=8.6cm]{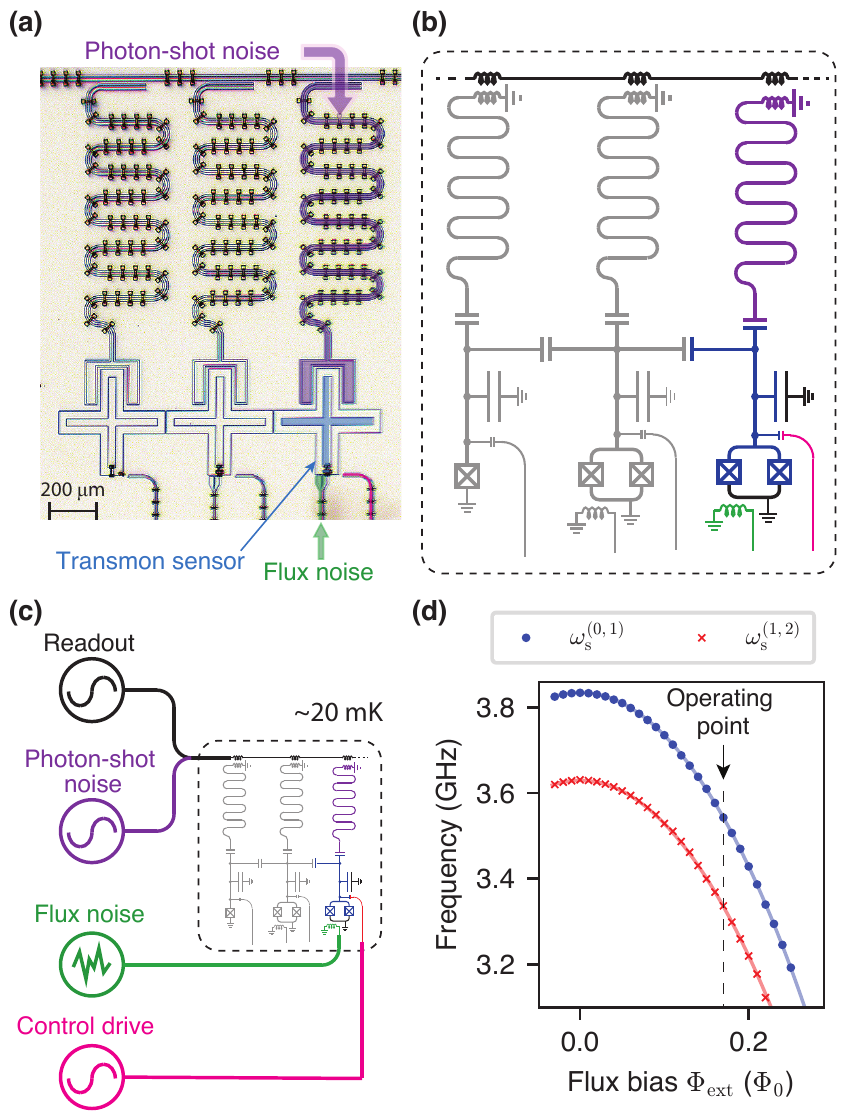}
    \caption{Device layout and simplifed experimental setup. 
    \textbf{(a)} Optical micrograph (false color) of the superconducting circuit comprising a flux-tunable transmon sensor (blue) to measure flux noise and photon shot noise applied via independent channels (green and purple, respectively). The transmon is controlled via a capacitively coupled drive line (magenta). 
    \textbf{(b)} Circuit schematic. The additional transmon qubits (grey) are far detuned from the frequency of the transmon sensor and can be neglected in this experiment. 
    \textbf{(c)} Simplified measurement schematic. Known, engineered flux noise and photon-shot noise is applied to the qubit. The control and readout lines are used to perform noise spectroscopy protocol and measure the results.
    \textbf{(d)} $|0\rangle$--$|1\rangle$ transition frequency (blue circles) and $|1\rangle$--$|2\rangle$ transition frequency (red {crosses}) of the transmon sensor as a function of the external flux bias $\Phi_{\mathrm{ext}}$. Solid lines correspond to simulations based on the circuit parameters (see {Supplementary Note}~1). The transmon sensor operates at a flux-sensitive point, $\Phi_{\mathrm{ext}}=\SI{0.17}{\Phi_{0}}$, See dashed black line.}
    \label{fig:Fig2}
\end{figure}

{
\subsection*{Experimental Validation}
}
We use the Xmon~\cite{Barend2013} variant of the superconducting flux-tunable transmon as a multi-level noise sensor. 
Our experimental test bed contains three transmon qubits, each of which is dispersively coupled to a coplanar-waveguide cavity for qubit state readout~\cite{Jeffrey2014,Sete2015}. 
In Fig.~\ref{fig:Fig2} (a) and (b), the rightmost transmon (blue) operates as a multi-level quantum sensor. 
The other transmons' modes are far-detuned from the sensor, such that their presence can be neglected 
(see {Supplementary Note}~1). 
In this work, we focus on two environmental noise channels that couple to the transmon sensor. 
One noise channel is formed by the inductive coupling of the sensor's SQUID loop to the fluctuating magnetic field in the qubit environment (flux noise). In this case, a fluctuating magnetic flux threading the SQUID loop results in the fluctuation of the qubit effective Josephson energy, thereby fluctuating the energy levels of the transmon sensor. 
The other noise source arises from photon number fluctuations in the readout resonator. In this case, photon-number fluctuations in the readout resonator cause a photon-number-dependent frequency shift of the energy levels of the sensor. 
Figure~\ref{fig:Fig2}(c) shows a reduced measurement schematic. 
We generate and apply a known level of engineered flux noise and coherent photon shot noise to the qubit, which we then use as a sensor to validate our protocol (see {Supplementary Note}~2).
We bias the transmon sensor at a flux-sensitive value $\Phi_{\mathrm{ext}}=\SI{0.17}{\Phi_0}$ (dashed line in Fig.~\ref{fig:Fig2}(d)). 
At this operating point, the energy relaxation times $T_1$ for $|0\rangle$--$|1\rangle$ and 
$|1\rangle$--$|2\rangle$ transitions are $\sim\SI{58}{\micro\second}$ and $\sim\SI{31}{\micro\second}$, respectively. 
Note that the energy relaxation time for the $|1\rangle$--$|2\rangle$ transition is approximately half that of the $|0\rangle$--$|1\rangle$ transition's relaxation time, which is expected for weakly anharmonic systems~\cite{Peterer2015}.

\begin{figure*}
    \begin{center}
    \includegraphics[width=17.78cm]{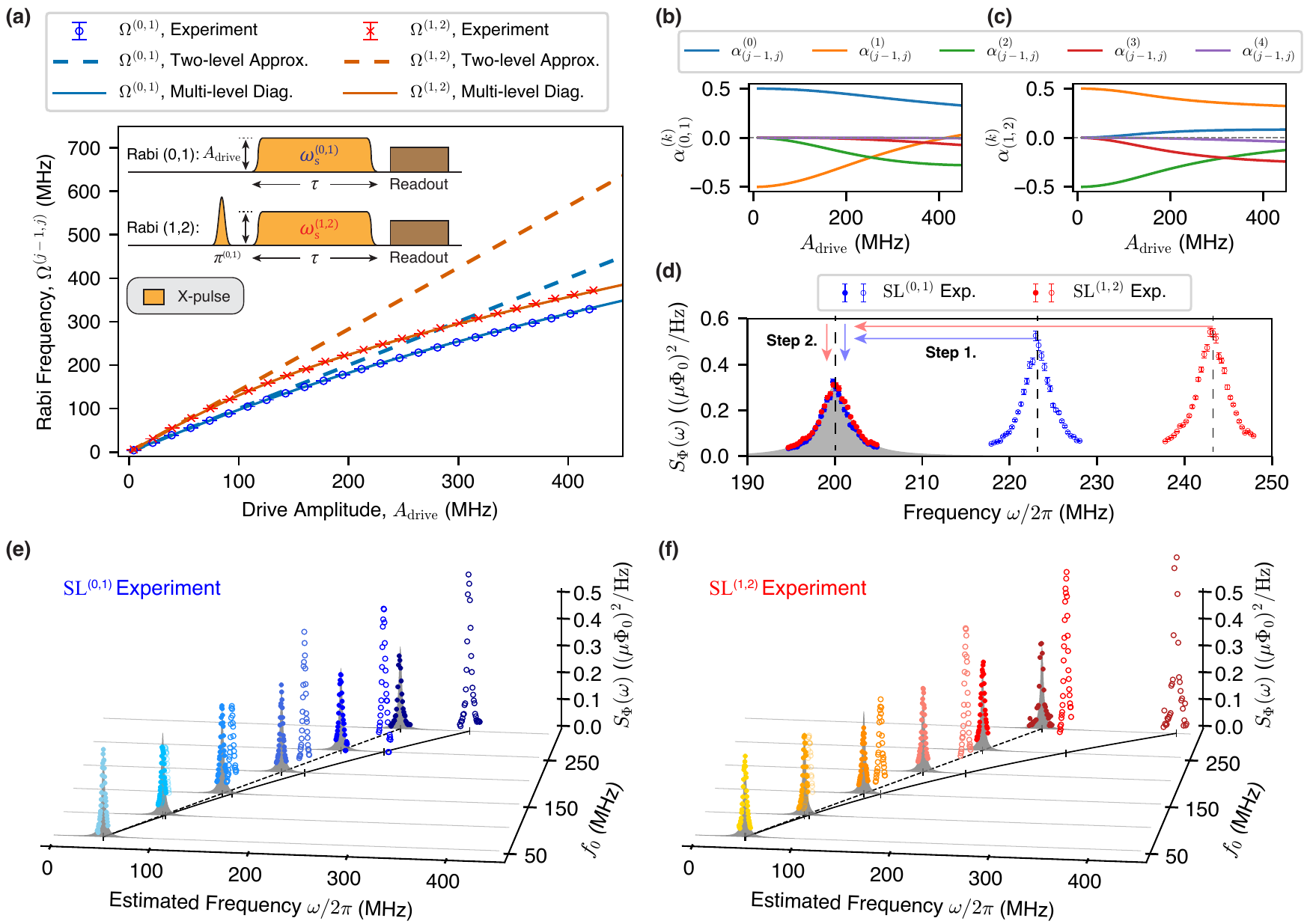}
    \end{center}
    \caption{Accurate spectral estimation of high-frequency noise. 
    \textbf{(a)} Rabi frequencies {$\Omega^{(j-1,j)}$} for $|0\rangle $--$|1\rangle$ (blue) and $|1\rangle $--$|2\rangle$ (red) transitions as a function of the drive amplitude $A_{\mathrm{drive}}$. Inset: pulse sequences used to measure the Rabi frequencies. 
    \textbf{(b)-(c)} Numerical calculations of noise participation ratios $\alpha^{(k)}_{(j-1,j)}$ as a function of drive amplitude $A_{\mathrm{drive}}$. 
    \textbf{(d)} Spectral estimation of the engineered flux noise (Lorentzian, centered at $ \SI{200}{\mega\hertz}$) via $\mathrm{SL}^{(0,1)}$ (blue) and $\mathrm{SL}^{(1,2)}$ (red) experiments. Two corrections are applied to the estimates under the two-level approximation {(hollow circles)}, and consist of shifting the frequency (step 1) and adjusting the magnitude (step 2) due to the multi-level dressing. Note that the corrected flux noise spectra for $\mathrm{SL}^{(0,1)}$ (blue circles) and $\mathrm{SL}^{(1,2)}$ (red circles) are in good agreement with the ideal flux noise PSD (grey filled). Error bars represent $\pm1$ standard deviations. %
    \textbf{(e)-(f)} Benchmarking the spectral estimation of engineered flux noise ranging from $f_0=\SI{50}{\mega\hertz}$ to $f_0=\SI{300}{\mega\hertz}$, where $f_0$ corresponds to the center frequency of engineered noise spectra. The different color shades of the PSD estimates correspond to engineered flux noise with different center frequencies. The agreement between the corrected experimental estimates (circles) and the ideal flux noise PSDs (grey filled) indicates that our protocol overcomes the spectral limit imposed by the sensor anharmonicity. 
    }
    \label{fig:Fig3}
\end{figure*} 

To test our protocol, we first demonstrate an accurate spectral reconstruction of engineered flux noise over a range of frequencies -- $\SI{50}{\mega\hertz}$ to $\SI{300}{\mega\hertz}$ -- that are smaller than, comparable to, and larger in magnitude than the transmon anharmonicity ($\omega^{(1,2)}_{\mathrm{s}} - \omega^{(0,1)}_{\mathrm{s}})/2\pi=-\SI{207.3}{\mega\hertz}$.
As with the standard spin-locking protocol, the transmon needs to be driven sufficiently strongly to form an energy splitting $\hbar\Omega^{(j-1,j)}$ between {a pair of spin-locked states} $\{ |+^{(j-1,j)}\rangle, |-^{(j-1,j)}\rangle \}$ at the measurement frequency of interest. 
However, when the splitting energy is comparable with or larger than the anharmonicity, the driven transmon can no longer be approximated as a two-level system, and the multi-level dressing that results must be carefully incorporated into the analysis to accurately reconstruct the PSD.

The first step in our noise spectrosopy demonstration is to measure the Rabi frequencies {$\Omega^{(j-1,j)}$} for the $|0\rangle $--$|1\rangle$ and $|1\rangle $--$|2\rangle$ transitions as a function of the drive amplitude $A_{\mathrm{drive}}$ [Fig.~\ref{fig:Fig3}(a)]. 
{
To determine $A_{\mathrm{drive}}$, we assume a linear dependence in the weak driving limit where Rabi frequency $<\SI{5}{\mega\hertz}$. From this linear dependence, we could extrapolate $A_{\mathrm{drive}}$ to the strong driving regime. 
}
For both the Rabi and the $\mathrm{SL}^{(j-1,j)}$ measurements, the rising and falling edges of the spin-locking drive envelope is Gaussian-shaped ($\propto\exp{(-t^2/2\sigma^2)}$) with $\sigma$ = $\SI{12}{\nano\second}$. 
For a given amplitude, the resulting Rabi frequency for the $|j-1\rangle $--$|j\rangle$ transition is equivalent to the level splitting $\Omega^{(j-1,j)}$ between the spin-locked states ($|+^{(j-1,j)}\rangle, |-^{(j-1,j)}\rangle$). 
However, recall that the measured Rabi frequencies $\Omega^{(j-1,j)}$ begin to deviate from the two-level system approximation ($\Omega^{(j-1,j)}=\lambda^{(j-1,j)}A_{\mathrm{drive}}$) as the drive amplitude is increased. 
The discrepancy $(\Omega^{(j-1,j)}-\lambda^{(j-1,j)}A_{\mathrm{drive}})$ is due to the multi-level dressing effect, the influence of other levels beyond the two-level approximation. 
Alternatively, one can also observe such frequency deviations by using pump-probe spectroscopy techniques (see {Supplementary Note}~3)~\cite{Baur2009,Braumuller15}. 
As such, the frequency shifts $(\Omega^{(j-1,j)} - \lambda^{(j-1,j)}A_{\mathrm{drive}})$ due to this multi-level dressing must be taken into account in order to obtain an accurate estimation of the flux noise spectra at frequencies comparable or larger than the anharmonicity.
In our experiments, we found that including up to the $4$'th excited state [solid curves in Fig.~\ref{fig:Fig3}(a)] was sufficient to obtain agreement between our numerical simulations and the experimentally observed frequency shifts. 

\begin{figure*}
    \begin{center}
    \includegraphics[width=17.78cm]{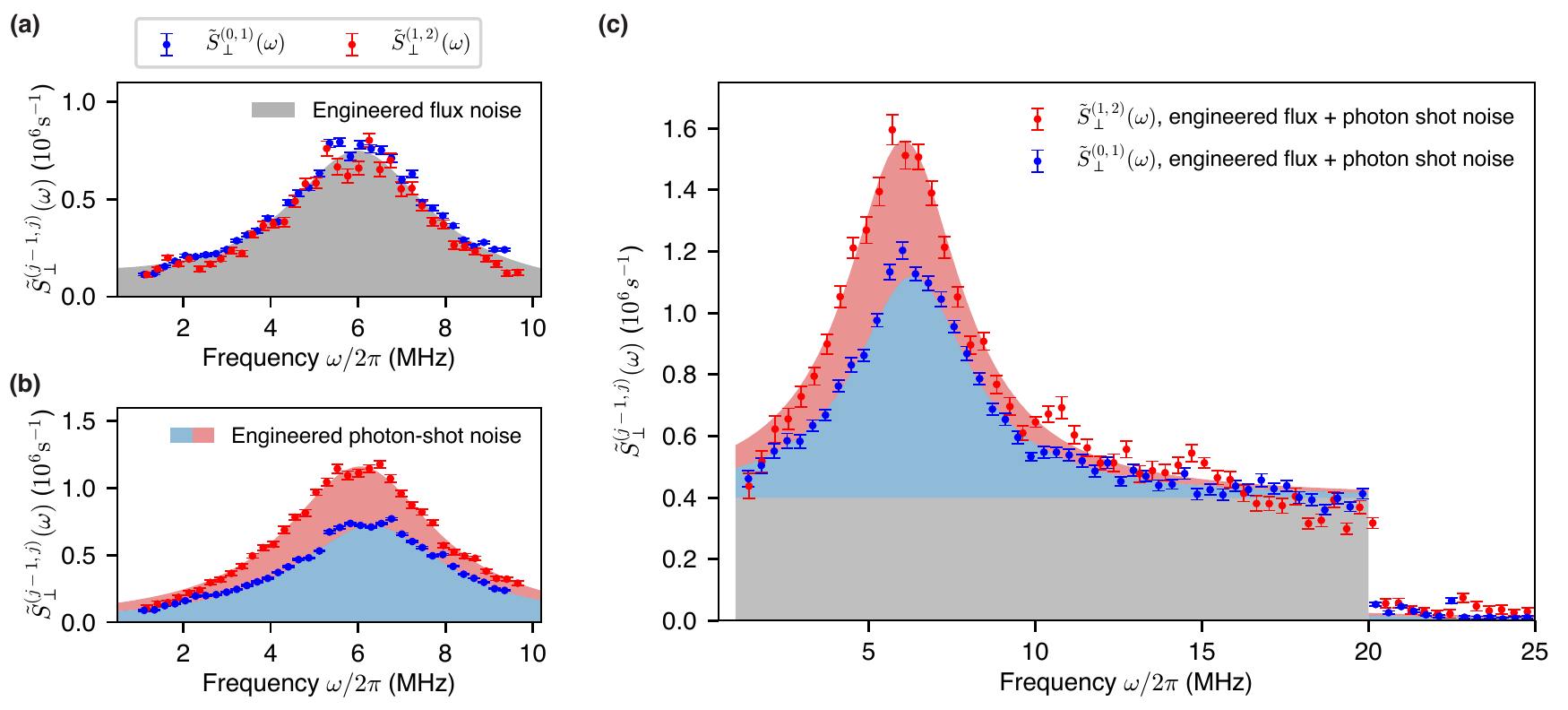}
    \end{center}
    \caption{Distinguishing the noise contributions from flux and photon shot noise. 
    \textbf{(a)} Transverse flux noise PSDs $\tilde{S}_{\Phi,\perp}^{(0,1)}(\omega)$ (blue) and $\tilde{S}_{\Phi,\perp}^{(1,2)}(\omega)$ (red) measured by performing $\mathrm{SL}^{(0,1)}$, $\mathrm{SL}^{(1,2)}$ for engineered Lorenztian flux noise (grey filled) centered at $\SI{6}{\mega\hertz}$. 
    \textbf{(b)} Transverse photon shot noise PSDs $\tilde{S}_{\bar{n},\perp}^{(0,1)}(\omega)$ (blue) and $\tilde{S}_{\bar{n},\perp}^{(1,2)}(\omega)$ (red) for engineered photon shot noise with detuning $\Delta/2\pi=\SI{6.05}{\mega\hertz}$ from the readout resonator. 
    \textbf{(c)} Total transverse noise PSDs $\tilde{S}_{\perp}^{(0,1)}(\omega)$ (blue) and $\tilde{S}_{\perp}^{(1,2)}(\omega)$ (red) for a mixture of engineered flux noise (grey-shaded box-car, $\SI{1}{\mega\hertz}$ to $\SI{20}{\mega\hertz}$), and coherent photon shot noise with detuning $\Delta/2\pi = \SI{6.05}{\mega\hertz}$ (blue- and red-shaded Lorentzians). Measuring the two-fold noise spectra $\tilde{S}_{\perp}^{(0,1)}(\omega)$ and $\tilde{S}_{\perp}^{(1,2)}(\omega)$ distinguishes the flux noise and photon shot noise contributions. Error bars represent $\pm1$ standard deviations.}
    \label{fig:Fig4}
\end{figure*}

Similarly, we must consider the noise participation of the peripheral bare states introduced through the multi-level dressing effect in order to obtain an accurate spectral esimation at high frequencies. 
To build intuition, we begin considering the low-frequency (small $A_{\mathrm{drive}}$) limit, where the longitudinal relaxation for $\mathrm{SL}^{(j-1,j)}$ is determined solely by dephasing noise that acts on $|j-1\rangle$ and $|j\rangle$ (Eq.~(\ref{eq:B_small_Omega})). 
Then, in the large $A_{\mathrm{drive}}$ limit, the flux noise acting on the peripheral levels also contributes to the longitudinal spin relaxation. 
Thus, we must use the noise participation ratios $\alpha_{(j-1,j)}^{(k)}$ for each energy level $k$, including the original two levels and the peripheral levels:
\begin{align}
    S_{\Phi}(\omega) = \tilde{S}^{(j-1,j)}_{\Phi,\perp}(\omega) \times \left(\sum_{k} \alpha_{(j-1,j)}^{(k)} \frac{\partial{\omega_{\mathrm{s}}^{(k)}}}{\partial \Phi_{\mathrm{ext}}}\right)^{-1} ,
\end{align} 
where $S_{\Phi}(\omega)$ denotes the power spectral density of the engineered flux noise at frequency $\omega/2\pi$, 
and $\partial \omega^{(k)}_{\mathrm{s}} / \partial \Phi_{\mathrm{ext}}$ denotes the flux noise sensitivity of the $k$'th level of the sensor. 
For our experiment, the values of $\alpha_{(j-1,j)}^{(k)}$ for $\mathrm{SL}^{(0,1)}$ and $\mathrm{SL}^{(1,2)}$ are numerically estimated and shown in Fig.~\ref{fig:Fig3}(b) and (c), respectively. 
We also numerically estimate $\partial \omega^{(k)}_{\mathrm{s}} / \partial \Phi_{\mathrm{ext}}$ for $k \in \{1,\cdots,4\}$ by solving the circuit Hamiltonian of the transmon sensor~(see {Supplementary Note}~1). 

We now reconstruct the spectrum of engineered Lorenzian-distributed flux noise centered at $\SI{200}{\mega\hertz}$, a frequency comparable to the sensor anharmonicity (see in Fig.~\ref{fig:Fig3}d).
For the sake of comparison, we first plot PSD estimates based on a two-level approximation {(hollow circles)}. 
The frequencies of these PSD estimates are shifted by $\lambda^{(j-1,j)}A_{\mathrm{drive}}-\Omega^{(j-1,j)}$ from the ideal flux noise spectra (grey shading). 
We would also conclude (erroneously) that the extracted flux noise PSD amplitude increases as the frequency increases when estimated using the two-level approximation.
In order to estimate the flux noise PSD accurately, the two corrections described above must be applied to the PSD estimates to account for the multi-level dressing effects: Step 1 -- a frequency shift; and Step 2 -- an amplitude adjustment. 
Upon applying these corrections, we successfully reconstruct the PSD estimates for the $\SI{200}{\mega\hertz}$ engineered flux noise for both $\mathrm{SL}^{(0,1)}$ and $\mathrm{SL}^{(1,2)}$ (markers lie on grey region, Fig.~\ref{fig:Fig3}d).

Using this approach, we benchmark the performance of $\mathrm{SL}^{(0,1)}$ and $\mathrm{SL}^{(1,2)}$ for a set of the Lorentzian-shaped engineered flux noise spectra which are centered at $f_0=$ 50, 100, 150, 200, 250, and $\SI{300}{\mega\hertz}$. 
Figure~\ref{fig:Fig3}(e) and (f) compare the ideal noise spectra (grey shading) with the corrected flux noise PSD estimates (circles sitting on the envelope of the grey regions and following a dashed line) measured by $\mathrm{SL}^{(0,1)}$ (blue shades) and $\mathrm{SL}^{(1,2)}$ (red shades), respectively, and with the uncorrected estimates (``x'' shapes following a solid line) that deviate in both the infered frequency and power. 
The different colors correspond to the different engineered flux noise spectra. 
The agreement between the corrected PSD estimates and the engineered noise PSDs clearly substantiates the idea that our protocol overcomes the anharmonicity limit of the noise sampling frequency by taking the multi-level dressing effect into account.

We now move on to distinguishing the noise contributions from both engineered flux and photon-shot noise by measuring $\mathrm{SL^{(0,1)}}$ and $\mathrm{SL^{(1,2)}}$. 
Importantly, both noise sources induce frequency fluctuations of the $|0\rangle$--$|1\rangle$ and $|1\rangle $--$|2\rangle$ transitions, but with a different and distinguishing relative noise power ($\tilde{S}^{(1,2)}_{\perp}(\omega)/\tilde{S}^{(0,1)}_{\perp}(\omega)$).  

In the case of flux noise, since the degree of transmon anharmonicity is independent of the external magnetic flux threading the transmon loop $\Phi_{\mathrm{ext}}$~\cite{Koch2007}, the flux-noise-induced frequency fluctuations of the $|0\rangle$--$|1\rangle$ and $|1\rangle $--$|2\rangle$ transitions are equal: $\partial \omega_{\mathrm{s}}^{(0,1)}/\partial \Phi_{\mathrm{ext}}=\partial \omega_{\mathrm{s}}^{(1,2)} /\partial \Phi_{\mathrm{ext}}$. 
Therefore, for low-frequency flux noise that causes dephasing, the relative noise power spectra of $\mathrm{SL}^{(1,2)}$ to $\mathrm{SL}^{(0,1)}$ is given as:   
\begin{align}
    \label{eq:S12_S01_flux_noise}
    \frac{\tilde{S}^{(1,2)}_{\Phi,\perp}(\omega)}{\tilde{S}^{(0,1)}_{\Phi,\perp}(\omega)} = 1,
\end{align}
where we have introduced the subscript $\Phi$ to indicate flux noise due to $\Phi_{\textrm{ext}}$.

In contrast, photon-shot noise induces frequency fluctuations for each level transition that scale with the corresponding effective dispersive strength $\chi^{(j-1,j)}$~\cite{Yan2018}.
The photon-number-dependent frequency shift due to photon shot noise affecting the $|j-1\rangle$--$|j\rangle$ transition is given as $\delta \omega^{(j-1,j)}_{\mathrm{s}} = 2\chi^{(j-1,j)} \bar{n}$, where $\bar{n}$ is the average residual photon number in the resonator. 
Hence, the relative noise power spectra of  $\mathrm{SL}^{(1,2)}$ to $\mathrm{SL}^{(0,1)}$ for photon shot noise is:
\begin{align}
    \label{eq:S12_S01_photon-shot_noise}
    \frac{\tilde{S}^{(1,2)}_{\bar{n},\perp}(\omega)}{\tilde{S}^{(0,1)}_{\bar{n},\perp}(\omega)} = \left(\frac{\chi^{(1,2)}}{\chi^{(0,1)}}\right)^2,
\end{align}
where we have introduced the subscript $\bar{n}$ to indicate photon shot noise.
This finding highlights the usefulness of measuring multiple noise spectra in order to deconvolve environmental noise processes.
We shall presume that these two independent sources of engineered noise -- flux noise and photon shot noise -- are the only two sources of transverse noise impacting the spin-locked spectrometers, and so we may define the total transverse noise power spectrum as $\tilde{S}^{(j-1,j)}_{\perp}(\omega)=\tilde{S}^{(j-1,j)}_{\Phi,\perp}(\omega)+\tilde{S}^{(j-1,j)}_{\bar{n},\perp}(\omega)$. 

We now demonstrate the identification and characterization of two independent noise sources by measuring the two-fold noise spectra $\tilde{S}_{\perp}^{(0,1)}(\omega)$ and $\tilde{S}_{\perp}^{(1,2)}(\omega)$.
To begin, in Fig.~\ref{fig:Fig4}(a), we present the experimentally extracted spectra $\tilde{S}_{\Phi,\perp}^{(0,1)}(\omega)$ (blue circles) and $\tilde{S}_{\Phi,\perp}^{(1,2)}(\omega)$ (red circles) for solely Lorentzian-shaped engineered flux noise centered at 6 MHz. 
The measured $\tilde{S}_{\Phi,\perp}^{(0,1)}(\omega)$ and $\tilde{S}_{\Phi,\perp}^{(1,2)}(\omega)$ are essentially identical, as one expects for transmon flux noise and consistent with Eq.~(\ref{eq:S12_S01_flux_noise}). 
Similarly, we also present the extracted $\tilde{S}_{\bar{n},\perp}^{(0,1)}(\omega)$ and $\tilde{S}_{\bar{n},\perp}^{(1,2)}(\omega)$ for solely engineered coherent photon-shot noise. 
The results are Lorentzian-shaped spectra centered at the frequency detuning $\Delta/2\pi\equiv(\omega_{\mathrm{r}}-\omega_{\mathrm{n}})/2\pi=6.05$ MHz between the readout resonator resonance frequency ($\omega_{\mathrm{r}}/2\pi$) and the applied coherent tone frequency ($\omega_{\mathrm{n}}/2\pi$) used to generate the shot noise~(Fig.~\ref{fig:Fig4}b)~\cite{Yan2018}. 
Contrary to the flux noise case, here the measured $\tilde{S}_{\bar{n},\perp}^{(0,1)}(\omega)$ and $\tilde{S}_{\bar{n},\perp}^{(1,2)}(\omega)$ have different magnitudes, with a measured ratio of $\tilde{S}_{\bar{n},\perp}^{(1,2)}(\omega) / \tilde{S}_{\bar{n},\perp}^{(0,1)}(\omega) \approx 1.61$ (see Eq.~\ref{eq:S12_S01_photon-shot_noise}), due to the differing values of $\chi^{(0,1)}$ and $\chi^{(1,2)}$. 
Next, we demonstrate the noise spectroscopy of a mixture of two engineered noise and identify their individual noise contributions. 
We inject flux noise ranging from 1 MHz to 20 MHz with a ``box-car'' envelope and, simultaneously, coherent photon shot noise from a coherent tone with a frequency that is detuned by $\Delta/2\pi=6.05$ MHz from the readout-resonator resonance frequency. 
Fig.~\ref{fig:Fig4}(c) presents the experimental data for the total transverse noise power spectra $\tilde{S}_{\perp}^{(0,1)}(\omega)$ (blue) and $\tilde{S}_{\perp}^{(1,2)}(\omega)$ (red), which include both flux noise and photon shot noise contributions. 
The known, engineered noise spectra for each transition is indicated by the grey box car (flux noise) and by the blue and red Lorentzians (shot noise) for the 0-1 and 1-2 transitions, respectively. 
Over the frequency domain where both flux and photon-shot noise are significant, 
(${\SI{3}{\mega\hertz}\leq\omega/2\pi\leq\SI{9}{\mega\hertz}}$), 
a distinction between $\tilde{S}_{\perp}^{(0,1)}(\omega)$ and $\tilde{S}_{\perp}^{(1,2)}(\omega)$ is clearly observed,
and the measured total noise spectral density is the sum of the flux noise and photon shot noise contributions, consistent with our assumption that these two noise sources are independent. 
In contrast, over the frequency domain where flux noise dominates (${\SI{15}{\mega\hertz}\leq\omega/2\pi\leq\SI{20}{\mega\hertz}}$), $\tilde{S}_{\perp}^{(0,1)}(\omega)$ and $\tilde{S}_{\perp}^{(1,2)}(\omega)$ are similar in magnitude and predominantly match the grey region. 
{Lastly, at 20 MHz, above which no external noise was applied, the data exhibit a discrete jump down to the 
sensitivity limit of the experiment (See {Supplementary Note}~6 for discussion on the sensitivity limit due to $T_1$ of the sensor).} 
This result indicates that we can distinguish the noise contributions from flux and photon shot noise by measuring the two-fold noise spectra $\tilde{S}_{\perp}^{(0,1)}(\omega)$ and $\tilde{S}_{\perp}^{(1,2)}(\omega)$. 
More generally, the independent extraction of unknown flux noise and photon shot noise would be performed by measuring $\tilde{S}_{\perp}^{(j-1,j)}(\omega)$ for a sufficient number of transitions $j-1,j$ and frequency ranges in order to back out the individual contributions (within certain and appropriate assumptions about the origin and type of noise). \\

\section*{Discussion}
In summary, we introduced and experimentally validated a noise spectroscopy protocol that utilizes multiple transitions of a qubit as a quantum sensor of its noise environment. 
By moving beyond the conventional two-level approximation, our approach overcomes the anharmonicity frequency limit of previous spin-locking approaches. 
We further show that measuring the noise spectra for multiple transitions enables one to distinguish certain noise sources, such as flux noise and photon shot noise, by leveraging the differing impact of those noise sources on the different transitions.
As an example, we measured the two-fold power spectra of dephasing noise acting on the $|0\rangle$--$|1\rangle$ and $|1\rangle$--$|2\rangle$ transitions of a transmon, and showed that our protocol can distinguish between externally applied, known, engineered noise contributions from flux noise and photon shot noise. 
We anticipate that applying this protocol to even higher level transitions ($j>2$) of a superconducting qubit sensor will enable one to distinguish other dephasing noise sources, such as charge noise~\cite{Christensen2019}. 

Although we mainly focus on the spin-locking based multi-level QNS throughout this work, extending the dynamic decoupling (D.D.) based noise spectroscopy protocols~\cite{Alvarez2011, Paz-Silva2014, Norris2016} to multi-level systems would also yield improved QNS performance {(see Supplementary Note 8 for a discussion of why we focus on the spin-locking based approaches rather than the D.D. based approaches throughout this work)}. Notably, the idea of discriminating noise sources by employing multiple level transitions as distinct spectrometers is immediately applicable to dynamic decoupling based approaches. 
In view of recent advances in optimal band-limited control~{\cite{Frey2020}}, we expect the implementation of dynamic decoupling based QNS using multi-level sensors will augment knowledge about noise sources in a manner similar to the spin-locking approach described here.

In this paper, we demonstrated our protocol by measuring engineered noise in the flux-tunable transmon sensor. We chose the operating point (flux bias) and operating frequency range (measured spectral range) of the sensor, such that it is dominantly affected by the engineered noise. However, the technique discussed here can be also applied to measure intrinsic noise of transmons such as $1/f$ flux noise~\cite{Oliver2013, Paladino2014, Braumuller2020}. Notably, by biasing the sensor at more flux-sensitive point, the sensitivity to flux noise can be further increased in order to detect intrinsic flux noise.

While we employ a flux-tunable transmon as a multi-level noise sensor, our methodology is portable to other anharmonic multi-level systems, such as the C-shunt flux qubit~\cite{Yan2016} and the fluxonium~\cite{Manucharyan2009, Nguyen2019}. Since the sensitivity of the qubit energies to various noise sources differ by qubit design, employing other superconducting qubits as multi-level noise sensors will enable us to explore various noise sources. We also envision the spin-locking QNS protocols – whether in a TLS approximation or a multi-level system – being used for other qubit modalities, such as quantum dot qubits or trapped ion qubits, as sensors of their local environments, such as their substrates or surface traps.

As detailed in {Supplementary Note}~6, the $T_1$ of the qubit can limit its noise sensitivity. However, as $T_1$ is improved through a combination of qubit design~\cite{Nguyen2019} and advanced materials~\cite{Place2020}, the sensitivity and utility of our approach also improves. Using diagnostic techniques such as the QNS protocol developed here to identify and characterize noise sources is an important step towards mitigating and eliminating them.

\section*{Data availability}
The data that support the findings of this study may be made available from the corresponding authors upon request and with the permission of the US Government sponsors who funded the work.

\section*{Code availability}
The code used for the analyses may be made available from the corresponding authors upon request and with the permission of the US Government sponsors who funded the work
\printbibliography 

@article{Preskill2018,
  doi = {10.22331/q-2018-08-06-79},
  %url = {https://doi.org/10.22331/q-2018-08-06-79},
  title = {Quantum {C}omputing in the {NISQ} era and beyond},
  author = {Preskill, John},
  journal = {{Quantum}},
  %issn = {2521-327X},
  publisher = {{Verein zur F{\"{o}}rderung des Open Access Publizierens in den Quantenwissenschaften}},
  volume = {2},
  pages = {79},
  year = {2018}
}

@article{Degen2017,
  title = {Quantum sensing},
  author = {Degen, C. L. and Reinhard, F. and Cappellaro, P.},
  journal = {Rev. Mod. Phys.},
  volume = {89},
  pages = {035002},
  numpages = {39},
  year = {2017},
  publisher = {American Physical Society},
  doi = {10.1103/RevModPhys.89.035002},
  %url = {https://link.aps.org/doi/10.1103/RevModPhys.89.035002}
}

@article{Paladino2014,
  title = {${1}/{f}$ noise: Implications for solid-state quantum information},
  author = {Paladino, E. and Galperin, Y. M. and Falci, G. and Altshuler, B. L.},
  journal = {Rev. Mod. Phys.},
  volume = {86},
  pages = {361--418},
  numpages = {58},
  year = {2014},
  publisher = {American Physical Society},
  doi = {10.1103/RevModPhys.86.361},
  %url = {https://link.aps.org/doi/10.1103/RevModPhys.86.361}
}

@article{Krantz2019,
   title={A quantum engineer`s guide to superconducting qubits},
   volume={6},
   %url={http://dx.doi.org/10.1063/1.5089550},
   DOI={10.1063/1.5089550},
   number={2},
   journal={Applied Physics Reviews},
   publisher={AIP Publishing},
   author={Krantz, P. and Kjaergaard, M. and Yan, F. and Orlando, T. P. and Gustavsson, S. and Oliver, W. D.},
   year={2019},
   pages={021318}
}

@incollection{Schoelkopf2002,
author = {Schoelkopf, R. J. and Clerk, A. A. and Girvin, S. M. and Lehnert, K. W. and Devoret, M. H.},
title = {{Qubits as spectrometers of quantum noise}},
editor = {Nazarov, Y. V.},
pages ={175--203},
booktitle={Quantum Noise in Mesoscopic Physics, NATO Science Series}, 
volume ={97},
publisher ={Springer, Dordrecht},
year = {2002}
}

@article{Alvarez2011,
author = {{\'{A}}lvarez, G. A. and Suter, D.},
doi = {10.1103/PhysRevLett.107.230501},
journal = {Phys. Rev. Lett.},
number = {23},
pages = {230501},
title = {{Measuring the spectrum of colored noise by dynamical decoupling}},
volume = {107},
year = {2011}
}

@article{Yuge2011,
author = {Yuge, T. and Sasaki, S. and Hirayama, Y.},
journal = {Phys. Rev. Lett.},
number = {17},
pages = {170504},
title = {{Measurement of the noise spectrum using a multiple-pulse sequence}},
volume = {107},
year = {2011} 
}

@article{Young2012,
author = {Young, K. C. and Whaley, K. B.},
journal = {Phys. Rev. A},
pages = {012314},
title = {{Qubits as spectrometers of dephasing noise}},
volume = {86},
year = {2012}
}

@article{Paz-Silva2014,
  title = {General Transfer-Function Approach to Noise Filtering in Open-Loop Quantum Control},
  author = {Paz-Silva, Gerardo A. and Viola, Lorenza},
  journal = {Phys. Rev. Lett.},
  volume = {113},
  pages = {250501},
  numpages = {5},
  year = {2014},
  publisher = {American Physical Society},
  doi = {10.1103/PhysRevLett.113.250501},
  %url = {https://link.aps.org/doi/10.1103/PhysRevLett.113.250501}
}

@article{Ithier2005,
  title = {Decoherence in a superconducting quantum bit circuit},
  author = {Ithier, G. and Collin, E. and Joyez, P. and Meeson, P. J. and Vion, D. and Esteve, D. and Chiarello, F. and Shnirman, A. and Makhlin, Y. and Schriefl, J. and Sch\"on, G.},
  journal = {Phys. Rev. B},
  volume = {72},
  pages = {134519},
  numpages = {22},
  year = {2005},
  publisher = {American Physical Society},
  doi = {10.1103/PhysRevB.72.134519},
  %url = {https://link.aps.org/doi/10.1103/PhysRevB.72.134519}
}

@article{Meriles2010,
author = {Meriles, C. A.  and Jiang, L.  and Goldstein, G.  and Hodges, J. S.  and Maze, J.  and Lukin, M. D.  and Cappellaro, P. },
title = {Imaging mesoscopic nuclear spin noise with a diamond magnetometer},
journal = {J. Chem. Phys.},
volume = {133},
pages = {124105},
year = {2010},
doi = {10.1063/1.3483676}
}

@article{Romach2015,
  title = {Spectroscopy of surface-induced noise using shallow spins in diamond},
  author = {Romach, Y. and M\"uller, C. and Unden, T. and Rogers, L. J. and Isoda, T. and Itoh, K. M. and Markham, M. and Stacey, A.   
  and Meijer, J. and Pezzagna, S. and Naydenov, B. and McGuinness, L. P. and Bar-Gill, N. and Jelezko, F.},
  journal = {Phys. Rev. Lett.},
  volume = {114},
  pages = {017601},
  year = {2015},
  doi = {10.1103/PhysRevLett.114.017601}
}

@article{Vandersypen05,
  title = {NMR techniques for quantum control and computation},
  author = {Vandersypen, L. M. K. and Chuang, I. L.},
  journal = {Rev. Mod. Phys.},
  volume = {76},
  pages = {1037--1069},
  numpages = {0},
  year = {2005},
  publisher = {American Physical Society},
  doi = {10.1103/RevModPhys.76.1037},
  %url = {https://link.aps.org/doi/10.1103/RevModPhys.76.1037}
}

@article{Carter2013,
  title = {Coherent manipulation of cold Rydberg atoms near the surface of an atom chip},
  author = {Carter, J. D. and Martin, J. D. D.},
  journal = {Phys. Rev. A},
  volume = {88},
  pages = {043429},
  numpages = {10},
  year = {2013},
  publisher = {American Physical Society},
  doi = {10.1103/PhysRevA.88.043429},
  %url = {https://link.aps.org/doi/10.1103/PhysRevA.88.043429}
}

@article{Bylander2011,
arxivId = {1101.4707},
author = {Bylander, J. and Gustavsson, S. and Yan, F. and Yoshihara, F. and Harrabi, K. and Fitch, G. and Cory, D. G. and 
Nakamura, Y. and Tsai, J. S. and Oliver, W. D.},
doi = {10.1038/nphys1994},
journal = {Nature Phys.},
number = {7},
pages = {565--570},
title = {{Noise spectroscopy through dynamical decoupling with a superconducting flux qubit}},
volume = {7},
year = {2011}
}

@article{Yan2013,
  title = {Rotating-frame relaxation as a noise spectrum analyser of a superconducting qubit undergoing driven evolution},
  author = {Yan, Fei and Gustavsson, Simon and Bylander, Jonas and Jin, Xiaoyue and Yoshihara, Fumiki and Cory, David G. and Nakamura, Yasunobu and Orlando, Terry P. and  Oliver, William D.},
  journal = {Nature Communications},
  volume = {4},
  pages = {2337},
  year = {2013},
  doi = {10.1038/ncomms3337},
  %url = {https://doi.org/10.1038/ncomms3337}
}

@article{Yoshihara2014,
author = {Yoshihara, F. and Nakamura, Y. and Yan, F. and Gustavsson, S. and Bylander, J. and Oliver, W. D. and Tsai, J. S.},
doi = {10.1103/PhysRevB.89.020503},
journal = {Phys. Rev. B},
number = {2},
pages = {020503(R)},
title = {{Flux qubit noise spectroscopy using {R}abi oscillations under strong driving conditions}},
volume = {89},
year = {2014}
}

@article{Quintana2017,
  title = {Observation of classical-quantum crossover of $1/f$ flux noise and its paramagnetic temperature dependence},
  author = {Quintana, C. M. and Chen, Yu and Sank, D. and Petukhov, A. G. and White, T. C. and Kafri, Dvir and Chiaro, B. and Megrant, A. 
  and Barends, R. and Campbell, B. and Chen, Z. and Dunsworth, A. and Fowler, A. G. and Graff, R. and Jeffrey, E. and Kelly, J. and 
  Lucero, E. and Mutus, J. Y. and Neeley, M. and Neill, C. and O'Malley, P. J. J. and Roushan, P. and Shabani, A. and Smelyanskiy, V. N. 
  and Vainsencher, A. and Wenner, J. and Neven, H. and Martinis, J. M.},
  journal = {Phys. Rev. Lett.},
  volume = {118},
  pages = {057702},
  year = {2017},
  doi = {10.1103/PhysRevLett.118.057702},
  %url = {https://link.aps.org/doi/10.1103/PhysRevLett.118.057702}
}

@article{Dial2013,
author = {Dial, O. E. and Shulman, M. D. and Harvey, S. P. and Bluhm, H. and Umansky, V. and Yacoby, A.},
journal = {Phys. Rev. Lett.},
number = {14},
pages = {146804},
title = {{Charge noise spectroscopy using coherent exchange oscillations in a singlet-triplet qubit}},
volume = {110},
year = {2013}
}

@article{Muhonen2014,
arxivId = {1402.7140},
author = {Muhonen, J. T. and Dehollain, J. P. and Laucht, A. and Hudson, F. E. and Kalra, R. and Sekiguchi, T. and Itoh, K. M. and Jamieson, D. N. and McCallum, J. C. and Dzurak, A. S. and Morello, A.},
doi = {10.1038/nnano.2014.211},
journal = {Nature Nanotech.},
pages = {986--991},
title = {{Storing quantum information for 30 seconds in a nanoelectronic device}},
volume = {9},
year = {2014}
}

@article{Morello2018,
  title = {Assessment of a silicon quantum dot spin qubit environment via noise spectroscopy},
  author = {Chan, K. W. and Huang, W. and Yang, C. H. and Hwang, J. C. C. and Hensen, B. and Tanttu, T. and Hudson, F. E. and Itoh, 
  K. M. and Laucht, A. and Morello, A. and Dzurak, A. S.},
  journal = {Phys. Rev. Applied},
  volume = {10},
  pages = {044017},
  year = {2018},
  doi = {10.1103/PhysRevApplied.10.044017}
}

@article{Tarucha2018,
  title = {A quantum-dot spin qubit with coherence limited by charge noise and fidelity higher than 99.9\%}, 
  author = {J. Yoneda and K. Takeda and T. Otsuka and T. Nakajima and M. R. Delbecq and G. Allison and T. Honda and T. Kodera and 
  Shunri Oda and Y. Hoshi and N. Usami and K. M. Itoh and S. Tarucha},  
  journal = {Nature Nanotech.},
  volume = {13},
  pages = {102--107},
  year = {2018}
 }

@article{Frey2017,
	Author = {V. M. Frey and S. Mavadia and L. M. Norris and W. de Ferranti and D. Lucarelli and L. Viola and M. J. Biercuk},
	Journal = {Nature Comms.},
	Pages = {2189},
	Title = {Application of optimal band-limited control protocols to quantum noise sensing},
	Volume = {8},
	Year = {2017}
}

@article{Norris2016,
  title = {Qubit Noise Spectroscopy for Non-Gaussian Dephasing Environments},
  author = {Norris, Leigh M. and Paz-Silva, Gerardo A. and Viola, Lorenza},
  journal = {Phys. Rev. Lett.},
  volume = {116},
  pages = {150503},
  numpages = {5},
  year = {2016},
  publisher = {American Physical Society},
  doi = {10.1103/PhysRevLett.116.150503},
  %%%url = {https://link.aps.org/doi/10.1103/PhysRevLett.116.150503}
}

@article{Sung2019,
abstract = {Accurate characterization of the noise influencing a quantum system of interest has far-reaching implications across quantum science, ranging from microscopic modeling of decoherence dynamics to noise-optimized quantum control. While the assumption that noise obeys Gaussian statistics is commonly employed, noise is generically non-Gaussian in nature. In particular, the Gaussian approximation breaks down whenever a qubit is strongly coupled to discrete noise sources or has a non-linear response to the environmental degrees of freedom. Thus, in order to both scrutinize the applicability of the Gaussian assumption and capture distinctive non-Gaussian signatures, a tool for characterizing non-Gaussian noise is essential. Here, we experimentally validate a quantum control protocol which, in addition to the spectrum, reconstructs the leading higher-order spectrum of engineered non-Gaussian dephasing noise using a superconducting qubit as a sensor. This first experimental demonstration of non-Gaussian noise spectroscopy represents a major step toward demonstrating a complete spectral estimation toolbox for quantum devices.},
author = {Sung, Youngkyu and Beaudoin, F{\'{e}}lix and Norris, Leigh M and Yan, Fei and Kim, David K and Qiu, Jack Y and von L{\"{u}}pke, Uwe and Yoder, Jonilyn L and Orlando, Terry P and Gustavsson, Simon and Viola, Lorenza and Oliver, William D},
doi = {10.1038/s41467-019-11699-4},
journal = {Nature Communications},
number = {1},
pages = {3715},
title = {{Non-Gaussian noise spectroscopy with a superconducting qubit sensor}},
%%url = {https://doi.org/10.1038/s41467-019-11699-4},
volume = {10},
year = {2019}
}

@article{Koch2007,
  title = {Charge-insensitive qubit design derived from the Cooper pair box},
  author = {Koch, Jens and Yu, Terri M. and Gambetta, Jay and Houck, A. A. and Schuster, D. I. and Majer, J. and Blais, Alexandre and Devoret, M. H. and Girvin, S. M. and Schoelkopf, R. J.},
  journal = {Phys. Rev. A},
  volume = {76},
  pages = {042319},
  numpages = {19},
  year = {2007},
  publisher = {American Physical Society},
  doi = {10.1103/PhysRevA.76.042319},
  %%url = {https://link.aps.org/doi/10.1103/PhysRevA.76.042319}
}

@article{Majer2007,
author = {Majer, J. and Chow, J. M. and Gambetta, J. M. and Koch, Jens and Johnson, B. R. and Schreier, J. A. and Frunzio, L. and Schuster, D. I. and Houck, A. A. and Wallraff, A. and Blais, A. and Devoret, M. H. and Girvin, S. M. and Schoelkopf, R. J.},
journal = {Nature},
volume = {449},
pages = {443-447},
title = {{Coupling superconducting qubits via a cavity bus}},
year = {2007}
}

@article{Larsen2015,
  title = {Semiconductor-Nanowire-Based Superconducting Qubit},
  author = {Larsen, T. W. and Petersson, K. D. and Kuemmeth, F. and Jespersen, T. S. and Krogstrup, P. and Nyg\aa{}rd, J. and Marcus, C. M.},
  journal = {Phys. Rev. Lett.},
  volume = {115},
  pages = {127001},
  numpages = {5},
  year = {2015},
  publisher = {American Physical Society},
  doi = {10.1103/PhysRevLett.115.127001},
  %%url = {https://link.aps.org/doi/10.1103/PhysRevLett.115.127001}
}

@article{Wang2019,
author = {Wang, Joel I.Jan and Rodan-Legrain, Daniel and Bretheau, Landry and Campbell, Daniel L. and Kannan, Bharath and Kim, David and Kjaergaard, Morten and Krantz, Philip and Samach, Gabriel O. and Yan, Fei and Yoder, Jonilyn L. and Watanabe, Kenji and Taniguchi, Takashi and Orlando, Terry P. and Gustavsson, Simon and Jarillo-Herrero, Pablo and Oliver, William D.},
journal = {Nature Nanotechnology},
volume = {14},
pages = {120-125},
title = {Coherent control of a hybrid superconducting circuit made with graphene-based van der Waals heterostructures},
year = {2019}
}

@article{Yan2016,
author = {Yan, F. and Gustavsson, S. and Kamal, A. and Birenbaum, J. and Sears, A. P. and Hover, D. and Gudmundsen, T. J. and Rosenberg, D. and Samach, G. and Weber, S. and Yoder, J. L. and Orlando, T. P. and Clarke, J. and Kerman, A. J. and Oliver, W. D.},
doi = {10.1038/ncomms12964},
journal = {Nature Communications},
pages = {12964},
pmid = {27808092},
title = {{The flux qubit revisited to enhance coherence and reproducibility}},
volume = {7},
year = {2016}
}

@article{Arute2019,
author = {Arute, Frank and Arya, Kunal and Babbush, Ryan and Bacon, Dave and Bardin, Joseph C and Barends, Rami and Biswas, Rupak and Boixo, Sergio and Brandao, Fernando G S L and Buell, David A and Burkett, Brian and Chen, Yu and Chen, Zijun and Chiaro, Ben and Collins, Roberto and Courtney, William and Dunsworth, Andrew and Farhi, Edward and Foxen, Brooks and Fowler, Austin and Gidney, Craig and Giustina, Marissa and Graff, Rob and Guerin, Keith and Habegger, Steve and Harrigan, Matthew P and Hartmann, Michael J and Ho, Alan and Hoffmann, Markus and Huang, Trent and Humble, Travis S and Isakov, Sergei V and Jeffrey, Evan and Jiang, Zhang and Kafri, Dvir and Kechedzhi, Kostyantyn and Kelly, Julian and Klimov, Paul V and Knysh, Sergey and Korotkov, Alexander and Kostritsa, Fedor and Landhuis, David and Lindmark, Mike and Lucero, Erik and Lyakh, Dmitry and Mandr{\`{a}}, Salvatore and McClean, Jarrod R and McEwen, Matthew and Megrant, Anthony and Mi, Xiao and Michielsen, Kristel and Mohseni, Masoud and Mutus, Josh and Naaman, Ofer and Neeley, Matthew and Neill, Charles and Niu, Murphy Yuezhen and Ostby, Eric and Petukhov, Andre and Platt, John C and Quintana, Chris and Rieffel, Eleanor G and Roushan, Pedram and Rubin, Nicholas C and Sank, Daniel and Satzinger, Kevin J and Smelyanskiy, Vadim and Sung, Kevin J and Trevithick, Matthew D and Vainsencher, Amit and Villalonga, Benjamin and White, Theodore and Yao, Z Jamie and Yeh, Ping and Zalcman, Adam and Neven, Hartmut and Martinis, John M},
doi = {10.1038/s41586-019-1666-5},
journal = {Nature},
number = {7779},
pages = {505--510},
title = {{Quantum supremacy using a programmable superconducting processor}},
%%url = {https://doi.org/10.1038/s41586-019-1666-5},
volume = {574},
year = {2019}
}

@article{Yan2018,
  title = {Distinguishing Coherent and Thermal Photon Noise in a Circuit Quantum Electrodynamical System},
  author = {Yan, Fei and Campbell, Dan and Krantz, Philip and Kjaergaard, Morten and Kim, David and Yoder, Jonilyn L. and Hover, David and Sears, Adam and Kerman, Andrew J. and Orlando, Terry P. and Gustavsson, Simon and Oliver, William D.},
  journal = {Phys. Rev. Lett.},
  volume = {120},
  pages = {260504},
  numpages = {6},
  year = {2018},
  publisher = {American Physical Society},
  doi = {10.1103/PhysRevLett.120.260504},
  %%url = {https://link.aps.org/doi/10.1103/PhysRevLett.120.260504}
}

@article{Luepke2020,
  title = {Two-Qubit Spectroscopy of Spatiotemporally Correlated Quantum Noise in Superconducting Qubits},
  author = {von L\"upke, Uwe and Beaudoin, F\'elix and Norris, Leigh M. and Sung, Youngkyu and Winik, Roni and Qiu, Jack Y. and Kjaergaard, Morten and Kim, David and Yoder, Jonilyn and Gustavsson, Simon and Viola, Lorenza and Oliver, William D.},
  journal = {PRX Quantum},
  volume = {1},
  pages = {010305},
  numpages = {23},
  year = {2020},
  publisher = {American Physical Society},
  doi = {10.1103/PRXQuantum.1.010305},
}

@article{Bishop2009,
abstract = {The exploration of the Jaynes–Cummings Hamiltonian in a circuit-QED system—where an ‘artificial atom' made of a superconducting circuit is strongly coupled to a microwave field—provides direct evidence for nonlinearities due to quantum mechanics on the level of single atoms and photons.},
author = {Bishop, Lev S and Chow, J M and Koch, Jens and Houck, A A and Devoret, M H and Thuneberg, E and Girvin, S M and Schoelkopf, R J},
doi = {10.1038/nphys1154},
journal = {Nature Physics},
number = {2},
pages = {105--109},
title = {{Nonlinear response of the vacuum Rabi resonance}},
%%url = {https://doi.org/10.1038/nphys1154},
volume = {5},
year = {2009}
}

@article{Braumuller15,
  title = {Multiphoton dressing of an anharmonic superconducting many-level quantum circuit},
  author = {Braum\"uller, Jochen and Cramer, Joel and Schl\"or, Steffen and Rotzinger, Hannes and Radtke, Lucas and Lukashenko, Alexander and Yang, Ping and Skacel, Sebastian T. and Probst, Sebastian and Marthaler, Michael and Guo, Lingzhen and Ustinov, Alexey V. and Weides, Martin},
  journal = {Phys. Rev. B},
  volume = {91},
  pages = {054523},
  numpages = {9},
  year = {2015},
  publisher = {American Physical Society},
  doi = {10.1103/PhysRevB.91.054523},
  %%url = {https://link.aps.org/doi/10.1103/PhysRevB.91.054523}
}

@article{Koshino2013,
  title = {Observation of the Three-State Dressed States in Circuit Quantum Electrodynamics},
  author = {Koshino, K. and Terai, H. and Inomata, K. and Yamamoto, T. and Qiu, W. and Wang, Z. and Nakamura, Y.},
  journal = {Phys. Rev. Lett.},
  volume = {110},
  pages = {263601},
  numpages = {5},
  year = {2013},
  publisher = {American Physical Society},
  doi = {10.1103/PhysRevLett.110.263601},
  %%url = {https://link.aps.org/doi/10.1103/PhysRevLett.110.263601}
}

@article{Barend2013,
  title = {Coherent Josephson Qubit Suitable for Scalable Quantum Integrated Circuits},
  author = {Barends, R. and Kelly, J. and Megrant, A. and Sank, D. and Jeffrey, E. and Chen, Y. and Yin, Y. and Chiaro, B. and Mutus, J. and Neill, C. and O'Malley, P. and Roushan, P. and Wenner, J. and White, T. C. and Cleland, A. N. and Martinis, John M.},
  journal = {Phys. Rev. Lett.},
  volume = {111},
  pages = {080502},
  numpages = {5},
  year = {2013},
  publisher = {American Physical Society},
  doi = {10.1103/PhysRevLett.111.080502},
  %%url = {https://link.aps.org/doi/10.1103/PhysRevLett.111.080502}
}

@article{Jeffrey2014,
  title = {Fast Accurate State Measurement with Superconducting Qubits},
  author = {Jeffrey, Evan and Sank, Daniel and Mutus, J. Y. and White, T. C. and Kelly, J. and Barends, R. and Chen, Y. and Chen, Z. and Chiaro, B. and Dunsworth, A. and Megrant, A. and O'Malley, P. J. J. and Neill, C. and Roushan, P. and Vainsencher, A. and Wenner, J. and Cleland, A. N. and Martinis, John M.},
  journal = {Phys. Rev. Lett.},
  volume = {112},
  pages = {190504},
  numpages = {5},
  year = {2014},
  publisher = {American Physical Society},
  doi = {10.1103/PhysRevLett.112.190504},
  %%url = {https://link.aps.org/doi/10.1103/PhysRevLett.112.190504}
}

@article{Sete2015,
  title = {Quantum theory of a bandpass Purcell filter for qubit readout},
  author = {Sete, Eyob A. and Martinis, John M. and Korotkov, Alexander N.},
  journal = {Phys. Rev. A},
  volume = {92},
  pages = {012325},
  numpages = {13},
  year = {2015},
  publisher = {American Physical Society},
  doi = {10.1103/PhysRevA.92.012325},
  %%url = {https://link.aps.org/doi/10.1103/PhysRevA.92.012325}
}

@article{Peterer2015,
  title = {Coherence and Decay of Higher Energy Levels of a Superconducting Transmon Qubit},
  author = {Peterer, Michael J. and Bader, Samuel J. and Jin, Xiaoyue and Yan, Fei and Kamal, Archana and Gudmundsen, Theodore J. and Leek, Peter J. and Orlando, Terry P. and Oliver, William D. and Gustavsson, Simon},
  journal = {Phys. Rev. Lett.},
  volume = {114},
  pages = {010501},
  numpages = {5},
  year = {2015},
  publisher = {American Physical Society},
  doi = {10.1103/PhysRevLett.114.010501},
  %%url = {https://link.aps.org/doi/10.1103/PhysRevLett.114.010501}
}

@article{Baur2009,
  title = {Measurement of Autler-Townes and Mollow Transitions in a Strongly Driven Superconducting Qubit},
  author = {Baur, M. and Filipp, S. and Bianchetti, R. and Fink, J. M. and G\"oppl, M. and Steffen, L. and Leek, P. J. and Blais, A. and Wallraff, A.},
  journal = {Phys. Rev. Lett.},
  volume = {102},
  pages = {243602},
  numpages = {4},
  year = {2009},
  publisher = {American Physical Society},
  doi = {10.1103/PhysRevLett.102.243602},
  %%url = {https://link.aps.org/doi/10.1103/PhysRevLett.102.243602}
}

@article{Christensen2019,
  title = {Anomalous charge noise in superconducting qubits},
  author = {Christensen, B. G. and Wilen, C. D. and Opremcak, A. and Nelson, J. and Schlenker, F. and Zimonick, C. H. and Faoro, L. and Ioffe, L. B. and Rosen, Y. J. and DuBois, J. L. and Plourde, B. L. T. and McDermott, R.},
  journal = {Phys. Rev. B},
  volume = {100},
  pages = {140503},
  numpages = {6},
  year = {2019},
  publisher = {American Physical Society},
  doi = {10.1103/PhysRevB.100.140503},
  %%url = {https://link.aps.org/doi/10.1103/PhysRevB.100.140503}
}

@article{Frey2020,
  title = {Simultaneous Spectral Estimation of Dephasing and Amplitude Noise on a Qubit Sensor via Optimally Band-Limited Control},
  author = {Frey, Virginia and Norris, Leigh M. and Viola, Lorenza and Biercuk, Michael J.},
  journal = {Phys. Rev. Applied},
  volume = {14},
  pages = {024021},
  numpages = {20},
  year = {2020},
  publisher = {American Physical Society},
  doi = {10.1103/PhysRevApplied.14.024021},
}

@article{Oliver2013,
author = {Oliver, W. D. and Welander, P. B.},
doi = {10.1557/mrs.2013.229},
journal = {MRS Bulletin},
number = {10},
pages = {816--825},
title = {{Materials in superconducting quantum bits}},
volume = {38},
year = {2013}
}

@article{Braumuller2020,
  title = {Characterizing and Optimizing Qubit Coherence Based on SQUID Geometry},
  author = {Braum\"uller, Jochen and Ding, Leon and Veps\"al\"ainen, Antti P. and Sung, Youngkyu and Kjaergaard, Morten and Menke, Tim and Winik, Roni and Kim, David and Niedzielski, Bethany M. and Melville, Alexander and Yoder, Jonilyn L. and Hirjibehedin, Cyrus F. and Orlando, Terry P. and Gustavsson, Simon and Oliver, William D.},
  journal = {Phys. Rev. Applied},
  volume = {13},
  pages = {054079},
  numpages = {11},
  year = {2020},
  publisher = {American Physical Society},
  doi = {10.1103/PhysRevApplied.13.054079},
  %%url = {https://link.aps.org/doi/10.1103/PhysRevApplied.13.054079}
}

@article {Manucharyan2009,
	author = {Manucharyan, Vladimir E. and Koch, Jens and Glazman, Leonid I. and Devoret, Michel H.},
	title = {Fluxonium: Single Cooper-Pair Circuit Free of Charge Offsets},
	volume = {326},
	number = {5949},
	pages = {113--116},
	year = {2009},
	doi = {10.1126/science.1175552},
	publisher = {American Association for the Advancement of Science},
	journal = {Science}
}

@article{Nguyen2019,
  title = {High-Coherence Fluxonium Qubit},
  author = {Nguyen, Long B. and Lin, Yen-Hsiang and Somoroff, Aaron and Mencia, Raymond and Grabon, Nicholas and Manucharyan, Vladimir E.},
  journal = {Phys. Rev. X},
  volume = {9},
  pages = {041041},
  numpages = {14},
  year = {2019},
  publisher = {American Physical Society},
  doi = {10.1103/PhysRevX.9.041041},
}

@Unpublished{Place2020,
%       title={New material platform for superconducting transmon qubits with coherence times exceeding 0.3 milliseconds}, 
%       author={Alex P. M. Place and Lila V. H. Rodgers and Pranav Mundada and Basil M. Smitham and Mattias Fitzpatrick and Zhaoqi Leng and Anjali Premkumar and Jacob Bryon and Sara Sussman and Guangming Cheng and Trisha Madhavan and Harshvardhan K. Babla and Berthold Jaeck and Andras Gyenis and Nan Yao and Robert J. Cava and Nathalie P. de Leon and Andrew A. Houck},
%       year={2020},
%       eprint={2003.00024},
%       archivePrefix={arXiv},
%       primaryClass={quant-ph}
% }
\section*{Acknowledgement}
It is a pleasure to thank F. Beaudoin, L. M. Norris, and L. Viola for insightful discussions, and M. Pulido for generous assistance. This research was funded by the U.S. Army Research Office grant No. W911NF-14-1-0682; and by the Department of Defense via MIT Lincoln Laboratory under Air Force Contract No. FA8721-05-C-0002. Y.S. acknowledges support from the Korea Foundation for Advanced Studies. The views and conclusions contained herein are those of the authors and should not be interpreted as necessarily representing the official policies or endorsements, either expressed or implied, of the U.S. Government.

{
\section*{Author contributions}
Y.S., J.B., A.V., and W.D.O conceived the project. Y.S., A.V., and S.G. performed the experiment and F.Y. and W.D.O. provided feedback. Y.S. developed the theoretical framework and carried out the numerical simulation with constructive feedback from A.V., J.B., F.Y., and W.D.O.. A.J.M., D.K.K., and J.L.Y. fabricated the device. J.B., J.W., M.K., R.W., A.B., and M.E.S. provided experimental assistance. T.P.O., S.G., and W.D.O. supervised the project. All authors contributed to the discussion of the results and the manuscript. 
}

\section*{Competing Interests}
Y.S., J.B., A.V., S.G., W.D.O, and Massachusetts Institute of Technology have filed a provisional US patent application related to multi-level quantum noise spectroscopy protocols. {All other authors have no competing interests.}

\clearpage 
\newpage 

\onecolumn

\section*{Supplementary Information}
\setcounter{section}{0}   
\renewcommand{\thefigure}{S\arabic{figure}}
\setcounter{figure}{0}
\renewcommand{\thetable}{S\arabic{table}}
\setcounter{table}{0}    

\section{Device parameters}
\label{SuppSec:DeviceParams}
In the experiments, we tune the transmon to an operating point where its frequency is highly sensitive to flux noise, see Fig.~2(d) and Supplementary Figure~1.
In addition to the sensor, there are two other transmon qubits in the same chip which are not used in the current experiment.
At the operating point, the nearest-neighboring (N.N.) transmon is far detuned from the transmon sensor, such that it can be neglected, while the sensor operates (qubit frequency detuning $\Delta_{\textrm{N.N.}}/2\pi\equiv (\omega_{\mathrm{N.N.}}^{(0,1)} - \omega_{\mathrm{s}}^{(0,1)})/2\pi$ = 1.6 GHz, effective coupling strength $g_{\mathrm{N.N.}}/2\pi \approx$ 15 MHz). Similarly, the next-nearest-neighboring (N.N.N.) transmon is also effectively isolated from the sensor, while it operates (qubit frequency detuning $\Delta_{\textrm{N.N.N.}}/2\pi\equiv (\omega_{\mathrm{N.N.N.}}^{(0,1)} - \omega_{\mathrm{s}}^{(0,1)})/2\pi$ = 230 MHz, effective coupling strength $g_{\mathrm{N.N.N.}}/2\pi \approx$ 1.4 MHz). The device parameters are summarized in Supplementary Table~\ref{table:device_params}.

\begin{table}[h!]
\caption{Device parameters.} 
\centering 
\begin{tabular}{c | c} 
\hline \hline 
Parameter &  Value \\ \hline 
Total Josephson energy of the two junctions $E_{\mathrm{J},\Sigma}$& 11.16 (GHz) \\
Junction asymmetry $d$& 0.0 \\
Capacitive energy $E_{\mathrm{c}}$& 181.5 (MHz) \\
Flux bias at the operating point $\Phi_{\mathrm{ext}}$& 0.170 ($\Phi_0$) \\
$|0\rangle$-$|1\rangle$ transition frequency at the operating point $\omega_{\mathrm{s}}^{(0,1)}/2\pi$& 3.5435 (GHz) \\ 
Relaxation time $T_1$ for the $|0\rangle$-$|1\rangle$ transition $1/\Gamma_1^{(0,1)}$& 58 ($\upmu$s) \\ 
$|1\rangle$-$|2\rangle$ transition frequency at the operating point $\omega_{\mathrm{s}}^{(1,2)}/2\pi$& 3.3362 (GHz) \\

Relaxation time $T_1$ for the $|1\rangle$-$|2\rangle$ transition $1/\Gamma_1^{(1,2)}$& 31 ($\upmu$s) \\ 
Readout resonator frequency $\omega_{\mathrm{r}}/2\pi$& 7.249 (GHz) \\
\footnote{We use the Purcell band-pass filter~[S3-S4] for fast readout.} Linewidth of the readout resonator  $\kappa_{r}/2\pi$& 4.18 (MHz) \\
Effective dispersive coupling strength for the  $|0\rangle$-$|1\rangle$ transition, $\chi^{(0,1)}/2\pi$ & 115 (kHz) \\
Effective dispersive coupling strength for the  $|1\rangle$-$|2\rangle$ transition, $\chi^{(1,2)}/2\pi$ & 146 (kHz) 
\end{tabular}
\label{table:device_params}
\end{table}

\begin{figure}[h!]
\centering
\includegraphics[width=17.8cm]{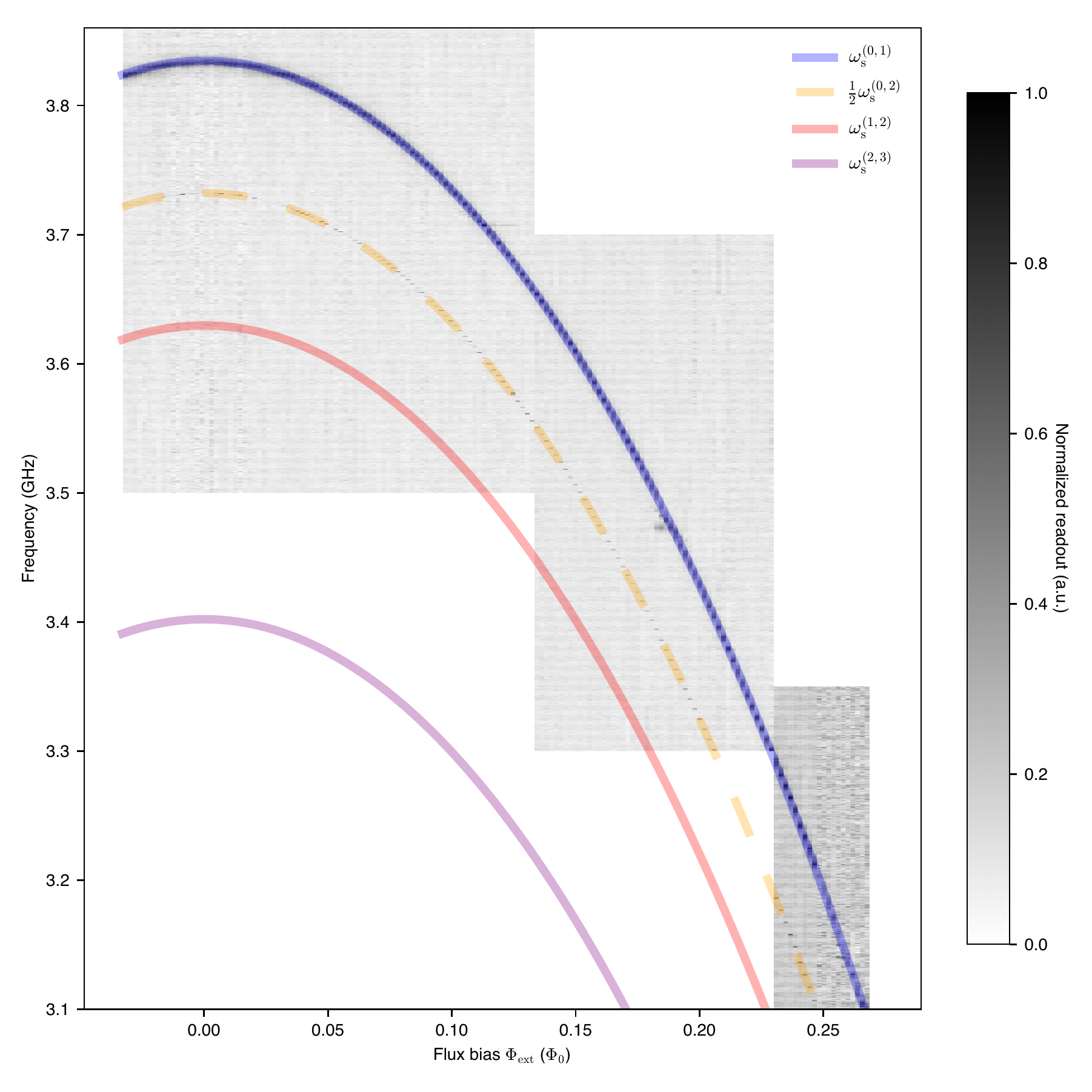}
\label{fig:SuppFig1}
\caption{Qubit spectroscopy as a function of the spectroscopy tone frequency and external magnetic flux threading the SQUID loop of the spectrometer. Solid and dashed curves are obtained by solving the eigen-energies of the circuit Hamiltonian based on the parameters summarized in Table.~\ref{table:device_params}. Note that the dashed orange curve corresponds to the two-photon ($|0\rangle$--$|2\rangle$) transition frequency.}
\end{figure}

\begin{figure}[h!]
\centering
\includegraphics[width=17.8cm]{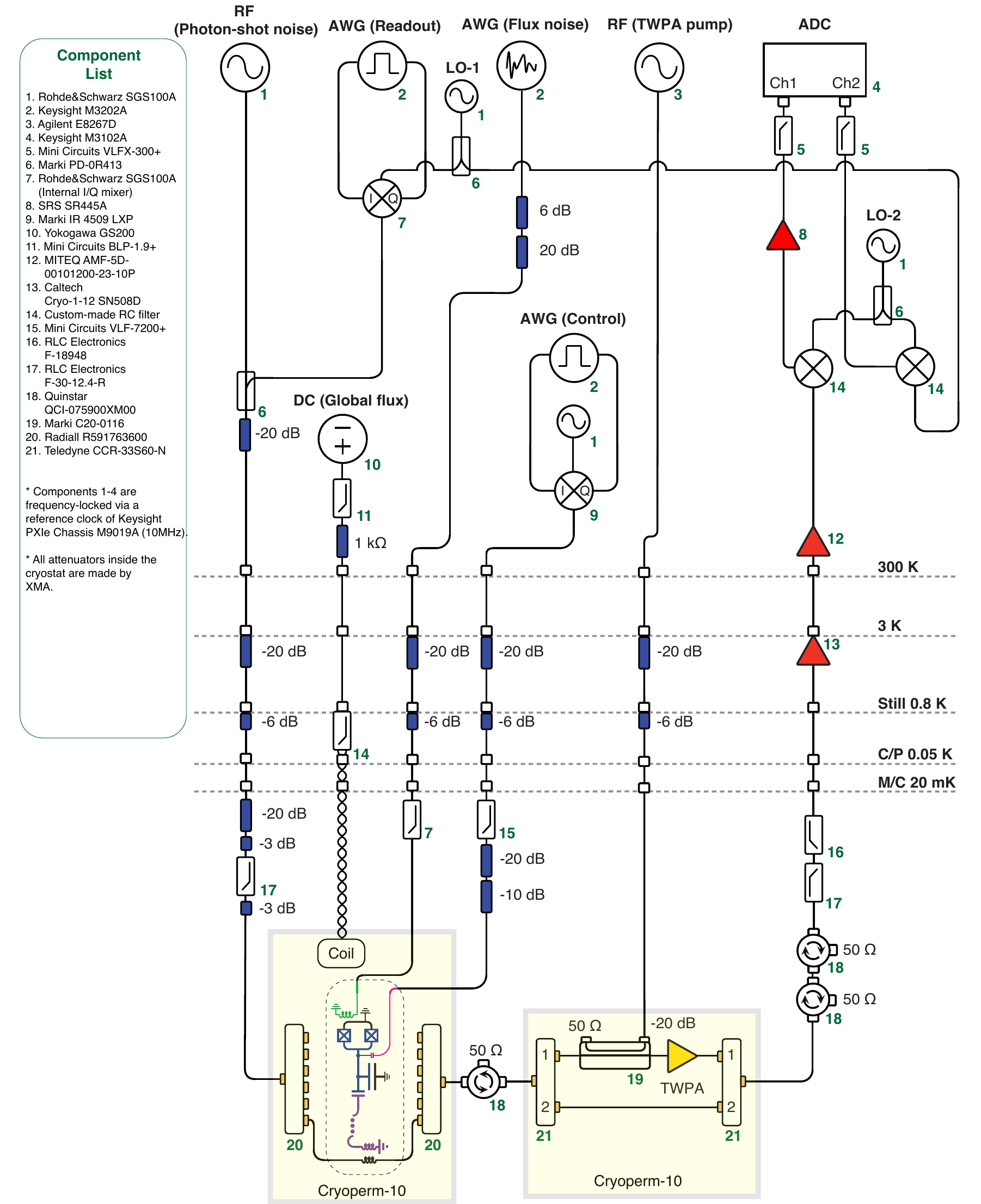}
\label{fig:SuppFig2}
\caption{Electronics and control wiring}
\end{figure}

\section{Measurement Setup}
\label{SuppSec:MeasurementSetup}
\subsection{Cryogenic setup}
The experiments were performed in a Leiden CF-450 dilution refrigerator with a base temperature of $\SI{20}{\milli\kelvin}$. The device was magnetically shielded with a superconducting can surrounded by a Cryoperm-10 cylinder. There are two lines for input and output; we apply microwave readout tone and measuring the transmission of sample.  All attenuators in the cryogenic samples are made by XMA and installed to remove excess thermal photons from higher-temperature stages. We pump the Josephson travelling wave parametric amplifier (JTWPA) to pre-amplify the readout signal at base tempearture~[S7]. 
To avoid any back-action of the pump-signal from TWPA, we added a microwave isolator between the  samples and the TWPA. On the RF output line, there is a high-electron mobility transistor (HEMT) amplifier (Cryo-1-12 SN508D) thermally connected to the $\SI{3}{\kelvin}$ stage. Two microwave isolators allow for the signal to pass through to the amplifier without being attenuated, while taking all the reflected noise off of the amplifier and dumping it in a $\SI{50}{\ohm}$ termination instead of reaching the sample. 

\subsection{Room temperature control}
Outside of the cryostat, we have all of the control electronics which allow us to apply microwave signals used for the readout and control of the transmon sensor.

All the signals are added using microwave power splitters (Marki PD0R413) used in reverse. Pulse envelopes of qubit control signals and readout signals are programmed in Labber software and then uploaded to arbitrary waveform generators (AWG Keysight M3202A). Subsequently, the pulses generated by AWGs are mixed with coherent tone from RF sources (Rohde and Schwarz SGS100A). All components for generating signals are frequency-locked using the \SI{10}{\mega\hertz} reference clock in the Keysight PXIe Chassis M9019A. A detailed schematic is given in Fig~S2. 

\subsection{Generation of engineered flux noise}
We generate engineered flux noise waveforms using the method described in Supplementary Material 3 of Ref.~[S5]. In all experiments presented in the main text, we consider the Lorentzian-shaped flux-noise PSD as follows:
\begin{align}
S_\Phi(\omega)=\frac{P_0}{2\pi\omega_c} \left(\frac{1}{1+\left[(\omega-\omega_0)/\omega_c\right]^2}+\frac{1}{1+\left[(\omega+\omega_0)/\omega_c\right]^2}\right),
\end{align}
where $P_0$ denotes the noise power, $\omega_0/2\pi=f_0$ denotes the center frequency of the noise, and $\omega_c/2\pi=\SI{2}{\mega\hertz}$ denotes the half-width at half-maximum (HWHM) of the Lorentzian curve. As described in [S5], we discretely sample the noise spectrum by taking harmonics separated by the fundamental frequency, $\SI{4}{\kilo\hertz}$. The noise spectrum is sampled with a high-frequency cutoff  ($\omega_0/2\pi+\SI{50}{\mega\hertz}$) and a low-frequency cutoff $\max(0,(\omega_0/2\pi-\SI{50}{\mega\hertz}))$. In each measurement, a new waveform is produced by an arbitrary waveform generator (AWG) and the total number of noise samples is 1,000. Each noise waveform has a duration of $\SI{100}{\micro\second}$.

\newpage
\clearpage\newpage
\section{Pump-probe spectroscopy}
\label{SuppSec:PumpProbeSpectroscopy}
In Fig.~3(a), we discussed the effect of multi-level dressing (the frequency shift, $\Omega-\lambda_j A_{\mathrm{drive}}$) by describing the deviation of measured Rabi frequency from the one expected in the two-level approximation. Here, we present experimental results of the pump-probe spectroscopy, which is an alternative approach to capture the effect of multi-level dressing.

Before describing the experimental results, we first present the dressed state picture for a driven multi-level system. We consider a multi-level transmon driven by an electromagnetic field, which is tuned to the frequency of the transmon's $|0\rangle$-$|1\rangle$ transition (Fig.~S1). Then, the transmon-photon system can be written as follows:
\begin{align}
    H(t) = H_{0} + H_{\mathrm{int}}(t),
\end{align}
where a time-independent Hamiltonian $H_0$ represents the sum of the energies of the transmon and the quantized mode of electromagnetic field (photon), and $H_{\mathrm{int}}(t)$ denotes the Hamiltonian describing the interaction between the transmon and the photon. We first choose a set of eigenstates of the interaction-free Hamiltonian $H_0$ as a basis. This basis corresponds to a tensor-product of the transmon states and photon states, $|j,n\rangle$ (Fig.~S3). In this basis, the transmon-photon system can be represented as multiple ladders of quantized energy levels. Now, we introduce the interaction between the transmon and photon (red double-headed arrow in Fig.~S4). In the case of a driven four-level transmon as illustrated in Fig.~S4, the four transmon-photon product states $|0, n\rangle, |1,n-1\rangle, |2,n-2\rangle$, and $|3,n-3\rangle$ can be grouped as ``the $n$-excitation manifold'' (grey-filled), each having the same number $n$ of excitations in total. The interaction between these product states leads to the formation of dressed states $ |+_n\rangle,  |-_n\rangle, |2'_n\rangle,$ and $|3'_n\rangle$; the dressed states correspond to the eigenstates of the Hamiltonian including the interaction. Note that single-photon transitions are allowed between the nearest-neighboring manifolds. Two-photon transitions are available between the next-nearest-neighboring manifolds.

\begin{figure}[h!]
\centering
\includegraphics[width=17.8cm]{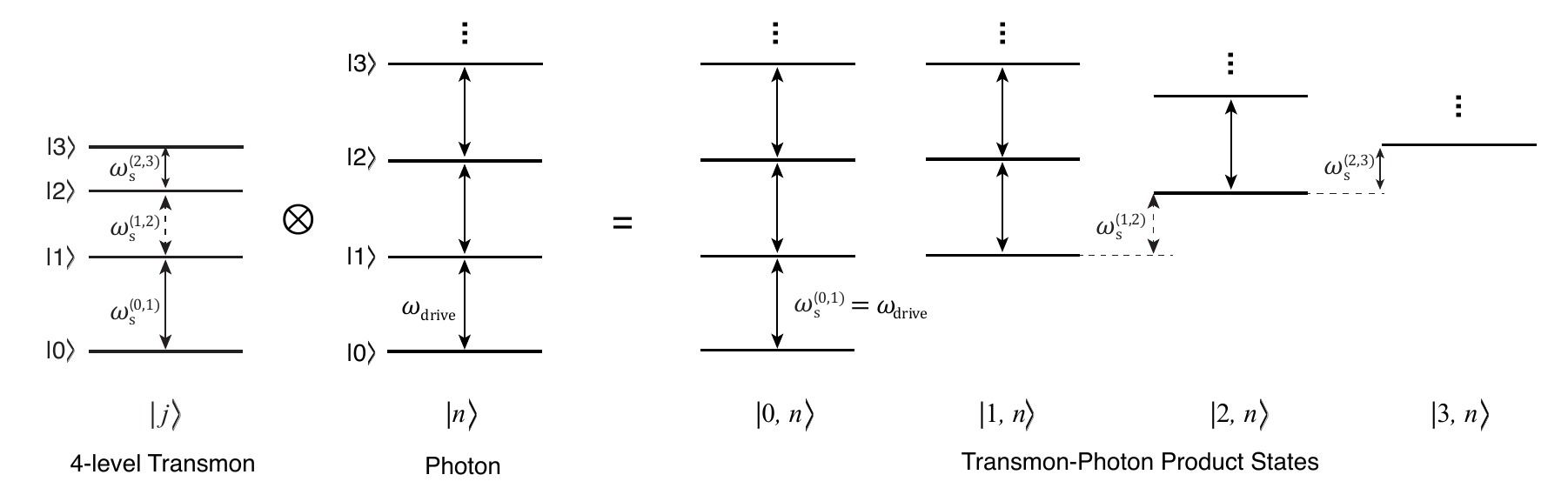}
\label{fig:SuppFig3}
\caption{Representation of (four-level-) transmon-photon product states. $j$ denotes the excitation level of the transmon and $n$ denotes the number of photons.}
\end{figure}

 \newpage
 
\begin{figure}[h!]
\centering
\includegraphics[width=17.8cm]{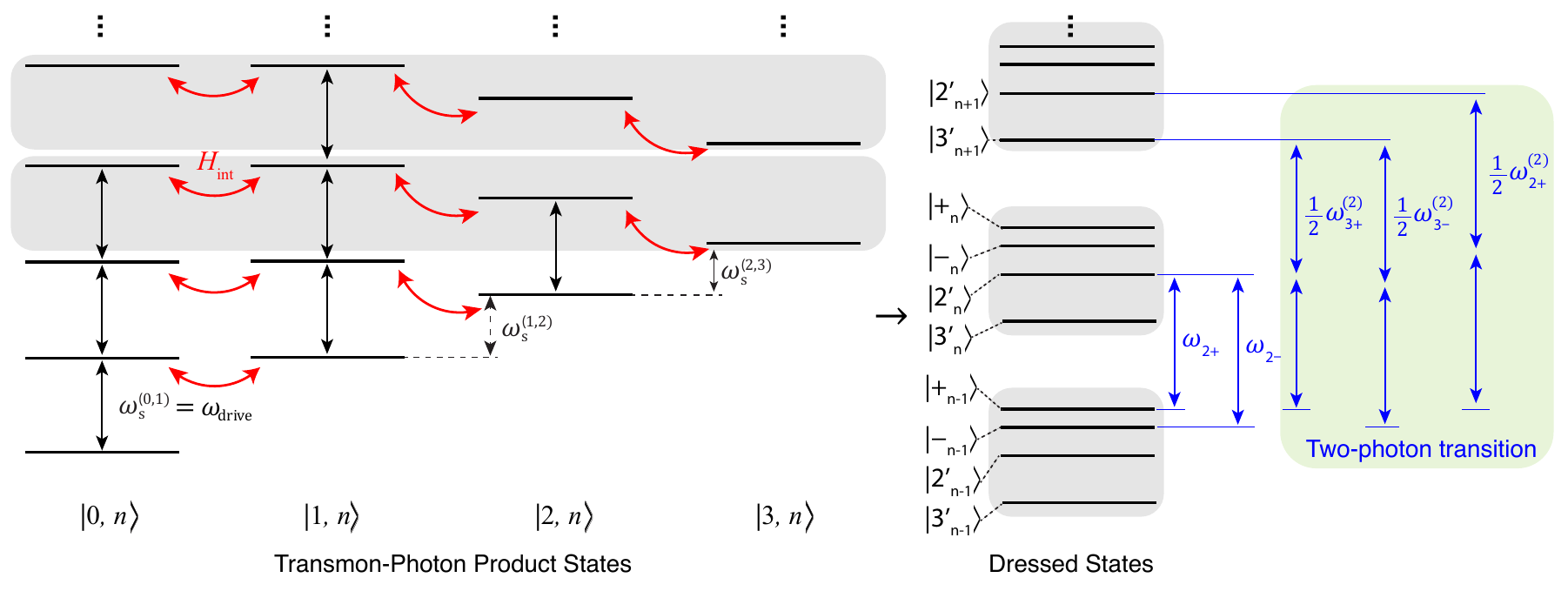}
\label{fig:SuppFig4}
\caption{Dressed state representation of the transmon-photon coupled system. The four product states $|0, n\rangle, |1,n-1\rangle, |2,n-2\rangle$, and $|3,n-3\rangle$ (grey filled) can be grouped as the $n$-excitation manifold. The interaction between the transmon and photon (red double-headed arrow, $H_{\mathrm{int}}$) leads to the formation of dressed states. Note that single-photon transitions are allowed between the nearest-neighboring manifolds. Two-photon transitions are available between the next-nearest-neighboring manifolds.}
\end{figure}

\begin{figure}[b!]
\centering
\includegraphics[width=12.9cm]{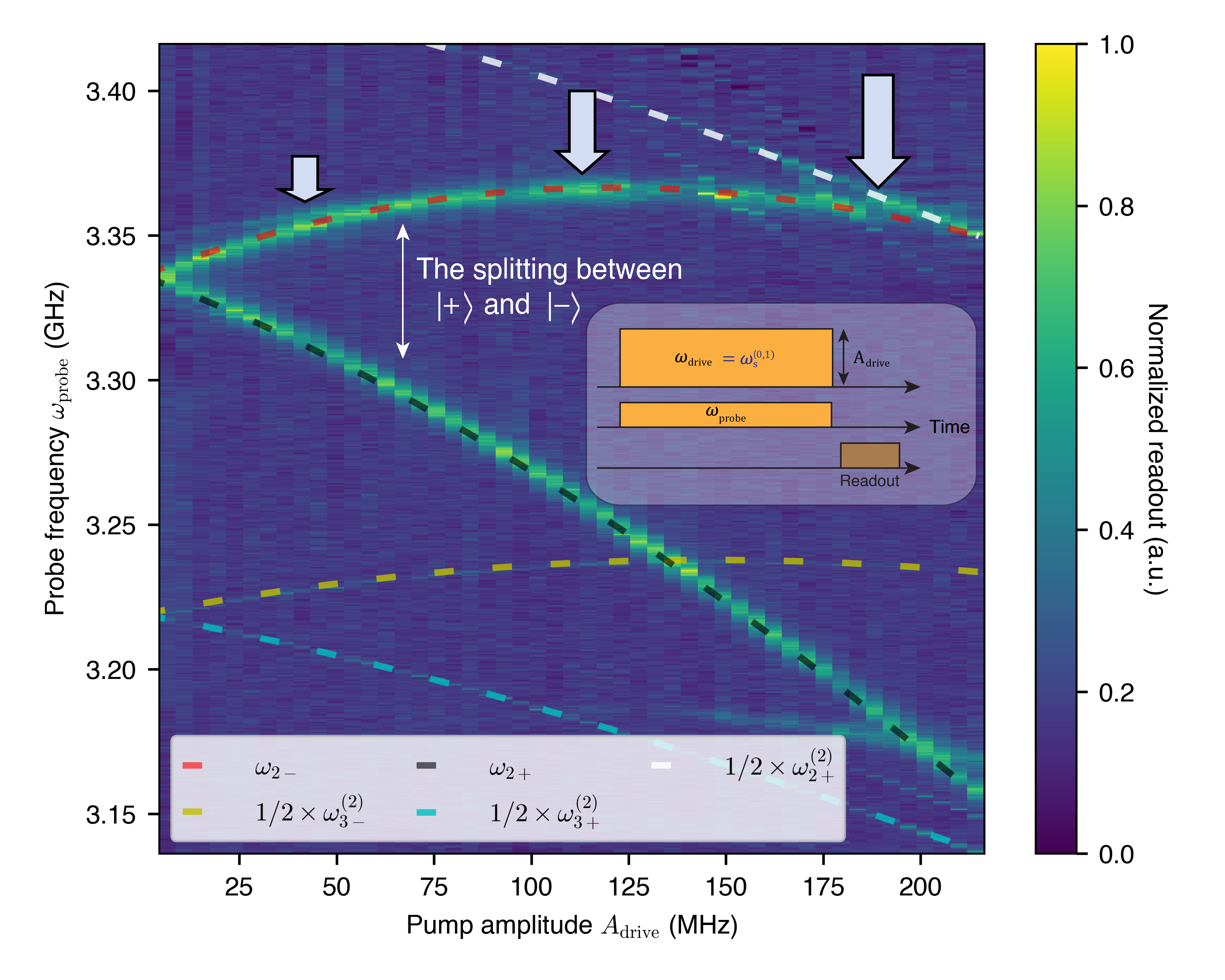}
\label{fig:SuppFig5}
\caption{Measurement of pump-probe spectroscopy. A strong pump tone is applied to the transmon sensor at the frequency of $|0\rangle$--$|1\rangle$ transition, $\omega_{\mathrm{s}}^{(0,1)}$, which dresses the transmon sensor. We measure single- and two-photon transition frequencies between the dressed states as a function of the pump amplitude ($A_{\mathrm{drive}}$) by measuring the absorption of a weak probe tone $(\omega_{\mathrm{probe}})$. The splitting between the dressed states $|+\rangle$ and $|-\rangle$ corresponds to the Vacuum Rabi splitting and is effectively pushed due the multi-level dressing effect. This effect becomes more significant as the pump amplitude $A_{\mathrm{drive}}$ increases. Dotted lines correspond to the simulation data based on the circuit parameters~(Table ~\ref{table:device_params}).}
\end{figure}
\newpage
In the pump-probe spectroscopy measurements, we drive the multi-level system with a strong pump tone at the frequency resonant with $|0\rangle$-$|1\rangle$ transition, which dresses the multi-level system. Then, we measure the level transitions between the dressed states by measuring the response of the system (absorption) to a weak probe tone. Fig.~S5 shows the measurement of transition frequencies as a function of the pump amplitude $A_{\mathrm{drive}}$ and compares with the simulation data. Note that the splitting between the dressed states $|+\rangle$ and $|-\rangle$ corresponds to the vacuum Rabi splitting $\Omega$. The vacuum Rabi splitting scales linearly with the drive amplitude (= pump amplitude) in the two-level approximation. However, in the case of a multi-level system, the splitting is effectively pushed due to the presence of higher transmon levels. The discrepancy between the vacuum Rabi splitting and the pump amplitude ($\Omega^{(j-1,j)}-A_{\mathrm{drive}}$) corresponds to the frequency shift due to the multi-level dressing, which is also measured in Fig.~3(a).  

\section{Derivation of the Effective Hamiltonian describing the $j$-th spin-locking noise spectrometer}
\label{SuppSec:Derivation_Heff}
As discussed in the main text, we consider an externally-driven $d$-level quantum sensor evolving under a noisy environment, which induces pure dephasing noise into the sensor. The Hamiltonian of the whole system $H(t)$ can be written as
\begin{align}
    H(t) = H_{\mathrm{S}}(t) + H_{\mathrm{SB}} + H_{\mathrm{B}},
\end{align}
where $H_{\mathrm{S}}(t)$ denotes the time-dependent Hamiltonian of the driven multi-level sensor, $H_{\mathrm{B}}$ denotes the bath Hamiltonian, and  $H_{\mathrm{SB}}$ denotes the sensor-bath interaction. Since we consider only pure-dephasing ($\sigma_z$-type) noise, the sensor-bath interaction can be written as
\begin{align}
    H_{\mathrm{SB}} = \hbar \sum_{j=1}^{d-1} B^{(j)} |j \rangle \langle j|, 
\end{align}
where $|j \rangle \langle j|$ is the projector for $j$-th level of the multi-level sensor and $B^{(j)}$ is the bath operator that longitudinally couples to the $j$-th level of the sensor~[S10]. Following the same notations used in the main text for Eq.~(1), the Hamiltonian of the driven multi-level sensor is given by
\begin{align}
    H_{\mathrm{S}}(t) =& \hbar \sum_{j=1}^{d-1} \Big [ \omega_{\mathrm{s}}^{(j)} |j \rangle \langle j| +  \lambda^{(j-1,j)} A_{\mathrm{drive}} \cos(\omega_{\mathrm{drive}}t) (\sigma_{+}^{(j-1,j)} + \sigma_{-}^{(j-1,j)}) \Big ].
\end{align}
To reiterate, the sensor eigenenergies are $\hbar \omega_{\mathrm{s}}^{(j)}$ with the ground state energy set to zero. The raising and lowering operators of the sensor are denoted by $\sigma_{+}^{(j-1,j)}\equiv|j\rangle \langle j-1|$ and  $\sigma_{-}^{(j-1,j)}\equiv|j-1\rangle \langle j|$, respectively.  
Here, we continuously drive the sensor with a continuous signal 
$A_{\mathrm{drive}} \cos(\omega_{\mathrm{drive}}t)$, where $A_{\mathrm{drive}}$ and $\omega_{\mathrm{drive}}$ correspond to the amplitude and the frequency of the driving field, respectively.
The parameter $\lambda^{(j-1,j)}$ represents the strength of the $|j-1\rangle $--$|j\rangle$ transition relative to the $|0\rangle$--$|1\rangle$ transition with  $\lambda^{(0,1)} \equiv 1$.

Now we move to the interaction picture with respect to the bath Hamiltonian $H_{\mathrm{B}}$. By introducing the time-dependent noise operator $B^{(j)}(t)\equiv e^{iH_{\mathrm{B}}t/\hbar}B^{(j)}e^{-iH_{\mathrm{B}}t/\hbar}$, the sensor-bath joint Hamiltonian can be written as (same as Eq.~(1) in the main text)
\begin{align}
    \tilde{H}(t) =& \hbar \sum_{j=1}^{d-1} \Big [ \left(\omega_{\mathrm{s}}^{(j)} +B^{(j)}(t) \right) |j \rangle \langle j| 
    +\lambda^{(j-1,j)} A_{\mathrm{drive}} \cos(\omega_{\mathrm{drive}}t) (\sigma_{+}^{(j-1,j)} + \sigma_{-}^{(j-1,j)}) \Big ],
\end{align}

Then, we move to the rotating frame, which rotates at the drive frequency $\omega_{\mathrm{drive}}$ with respect to the longitudinal axis of the sensor. To move to the rotating frame, we define the unitary operator $U_{\mathrm{R}}(t)$ as
\begin{align}
    U_{\mathrm{R}}(t) \equiv \sum_{j=1}^{d-1} \exp{(-ij\omega_{\textrm{drive}}t) |j \rangle \langle j|},
\end{align}
which determines the transformed Hamiltonian as follows:
\begin{align}
    \tilde{H}_{\mathrm{R}} (t) = U_{\mathrm{R}}^{\dagger}(t) \tilde{H}(t) U_{\mathrm{R}}(t) + i \dot{U}_{\mathrm{R}}^{\dagger}(t)U_{\mathrm{R}}(t).
\end{align}
Assuming that $\omega_{\mathrm{drive}}$ is larger then any other rate or frequency in this frame, we can perform the rotating wave approximation (RWA), which leads to the Hamiltonian~[S7]
\begin{align}
  &\tilde{H}_{\mathrm{RWA}}(t) = \hbar \sum_{j=1}^{d-1} \Bigg[ \Big[\left( \omega_{\mathrm{s}}^{(j)} - j\omega_{\mathrm{drive}} \right) + B^{(j)}(t) \Big] |j\rangle \langle j| 
  +\lambda^{(j-1,j)} \left[ \frac{A_{\mathrm{drive}}}{2} (\sigma^{(j-1,j)}_+ + \sigma^{(j-1,j)}_-)  \right] \Bigg]
\nonumber \\
  &=\hbar \left[
    \begin{pmatrix}
    0 & A_{\mathrm{drive}}/2 & 0 & \cdots & 0\\
    A_{\mathrm{drive}}/2 & \omega_{\mathrm{s}}^{(1)}-\omega_{\mathrm{drive}} & \lambda^{(1,2)}A_{\mathrm{drive}}/2 & \cdots & 0 \\
    0 & \lambda^{(1,2)}A_{\mathrm{drive}}/2 & \omega_{\mathrm{s}}^{(2)}-2\omega_{\mathrm{drive}} & \cdots & 0\\
    \vdots & \vdots & \vdots & \ddots & \vdots \\
    0 & 0 & 0 & \cdots & \omega_{\mathrm{s}}^{(d-1)}-(d-1)\omega_{\mathrm{drive}}
    \end{pmatrix}
    + 
    \begin{pmatrix}
    0 & 0 & 0 & \cdots & 0\\
    0 & B^{(1)}(t) & 0 & \cdots & 0 \\
    0 & 0 & B^{(2)}(t) & \cdots & 0\\
    \vdots & \vdots & \vdots & \ddots & \vdots \\
    0 & 0 & 0 & \cdots & B^{(d-1)}(t)
    \end{pmatrix}
    \right]
    \nonumber \\
&\equiv \tilde{H}_{\mathrm{S,RWA}} + \hbar \sum_{j=1}^{d-1}  B^{(j)}(t)  |j \rangle \langle j|,
\end{align} 
where we have introduced the system Hamiltonian ${\tilde{H}_{\textrm{S,RWA}} \equiv \hbar \sum_{j=1} \Big[\left( \omega_{\mathrm{s}}^{(j)} - j\omega_{\mathrm{drive}} \right) +\frac{1}{2} \lambda^{(j-1,j)} A_{\mathrm{drive}} \left(  \sigma^{(j-1,j)}_+ + \sigma^{(j-1,j)}_- \right) \Big]}$, which does not include the dephasing term.

Next, we find a change-of-basis matrix $V$ that diagonalizes the system Hamiltonian $\tilde{H}_{\mathrm{S,RWA}}$, 
\begin{align}
    {V}^{\dagger}  \tilde{H}_{\mathrm{S,RWA}}{V}=&
    \hbar {V}^{\dagger} 
    \begin{pmatrix}
    0 & A_{\mathrm{drive}}/2 & 0 & \cdots & 0\\
    A_{\mathrm{drive}}/2 & \omega_{\mathrm{s}}^{(1)}-\omega_{\mathrm{drive}} & \lambda^{(1,2)}A_{\mathrm{drive}}/2 & \cdots & 0 \\
    0 & \lambda^{(1,2)}A_{\mathrm{drive}}/2 & \omega_{\mathrm{s}}^{(2)}-2\omega_{\mathrm{drive}} & \cdots & 0\\
    \vdots & \vdots & \vdots & \ddots & \vdots \\
    0 & 0 & 0 & \cdots & \omega_{\mathrm{s}}^{(d-1)}-(d-1)\omega_{\mathrm{drive}}
    \end{pmatrix}
    {V}
    \nonumber \\
    =&
    \hbar 
    \begin{pmatrix}
    E^{(0)}& 0 & 0 & \cdots & 0\\
    0 & E^{(1)} & 0 & \cdots & 0 \\
    0 & 0 & E^{(2)} & \cdots & 0\\
    \vdots & \vdots & \vdots & \ddots & \vdots \\
    0 & 0 & 0 & \cdots & E^{(d-1)} 
    \end{pmatrix},
\end{align}
where the eigenenergies $\hbar E^{(j)}$ of the $j$-th dressed states lie on its diagonal entries. Note that the matrix $V$ may not diagonalize the full Hamiltonian $\tilde{H}_{\textrm{RWA}}$. 
The frame where the system Hamiltonian $\tilde{H}_{\rm S,RWA}$ is diagnoalized is referred to as the spin locking frame. The full Hamiltonian in the spin locking frame is
\begin{align}
    \tilde{H}_{\mathrm{SL}} =&  {V}^{\dagger}  \tilde{H}_{\mathrm{RWA}}(t) {V} = {V}^{\dagger}  \tilde{H}_{\mathrm{S,RWA}}{V} + {V}^{\dagger} \left( \hbar \sum_{j=1}^{d-1}  B^{(j)}(t)  |j \rangle \langle j| \right) {V} \nonumber \\ 
    =& \hbar \left[ \mathrm{diag}( E^{(0)},  E^{(1)}, \cdots  E^{(d-1)}) + 
     \sum_{j=1}^{d-1} B^{(j)}(t) {V}^{\dagger} |j \rangle \langle j|{V} \right].
\end{align}

Since the matrix $V$ is not diagonal in general, the above equation clearly shows that longitudinal noise in the lab frame for a multi-level system leads to both transverse and longitudinal noise in the  spin-locking frame. Note that this is not the case for a two-level system. In that case, longitudinal noise in the laboratory frame is fully transduced to transverse noise in the spin-locking frame.

For $\omega_{\mathrm{drive}}=\omega_{\mathrm{s}}^{(j)}-\omega_{\mathrm{s}}^{(j-1)}$, the relevant pair of the dressed states are $|+^{(j-1,j)}\rangle, |-^{(j-1,j)}\rangle$, with energy splitting $\hbar\Omega^{(j-1,j)}=(E^{(j)}-E^{(j-1)})$. %
This pair of dressed states spans an effective two-level subspace which forms the $j$-th spin-locking spectrometer. In order to describe the dynamics of the $j$-th spectrometer, we truncate the $d$-dimensional Hilbert space of the multi-level system to its two-dimensional subspace.
The truncated Hamiltonian $\tilde{H}_{\mathrm{SL}}^{(j-1,j)}$ is given by
\begin{align}
    \label{eq:trunc_H_spectrometer}
    \tilde{H}_{\mathrm{SL}}^{(j-1,j)} = \hbar
    \left[
    \begin{pmatrix}
    E^{(j-1)} & 0\\
    0 & E^{(j)}
    \end{pmatrix}
    +
    \begin{pmatrix}
    \sum_{k=1}^{d-1} B^{(k)}(t) \langle j-1 | V^{\dagger} |k \rangle \langle k| V |j-1\rangle & \sum_{k=1}^{d-1} B^{(k)}(t) \langle j-1 | V^{\dagger} |k \rangle \langle k| V |j\rangle \\
    \sum_{k=1}^{d-1} B^{(k)}(t) \langle j-1 | V^{\dagger} |k \rangle \langle k| V |j\rangle & \sum_{k=1}^{d-1} B^{(k)}(t) \langle j | V^{\dagger} |k \rangle \langle k| V |j\rangle
    \end{pmatrix}
    \right].
\end{align}
Note that the behavior of the $j$-th spectrometer is accurately described by Eq.~\eqref{eq:trunc_H_spectrometer} only if leakage out of this subspace can be neglected.

We evaluate the leakage rate from $|+\rangle^{(j-1,j)}$ (with eigenenergy $E^{(j)}$) to the energetically nearest dressed state outside of the subspace (with eigenenergy $E^{(j+1)}$) by applying Fermi's golden rule~[S11], and analogously for $|-\rangle^{(j-1,j)}$. Note that here we consider leakage caused by dephasing noise, characterized by $B(t)$. The corresponding transition (leakage) rate $\Gamma_{\mathrm{leak}}^{(j)}$ can be expressed in terms of the noise spectral density $S(\omega)\equiv \int_{-\infty}^{\infty}\diff{\tau} E^{-i\omega \tau} \langle {B}(\tau){B}(0)\rangle$ as follows:
\begin{align}
    \Gamma_{\mathrm{leak}}^{(j)} = \frac{A^2}{\hbar^2} S(\omega = E^{(j+1)} - E^{(j)})
\end{align}
$A=\sum_{k=1}^{d-1} \langle j+1 | V^{\dagger} |k \rangle \langle k| V |j\rangle$ denotes the matrix element for the corresponding transition. In the limit of a weak spin-locking drive as used in our experiment, the matrix element quantifying leakage out of the subspace is negligible compared to the matrix element between the spin-locked states, such that leakage is suppressed. In addition, the transition frequency to levels outside the spectrometer subspace is larger than the frequency splitting $\Omega^{(j-1,j)}$. For a noise PSD that decays for higher frequencies (e.g. $1/f$ noise), these high frequency transitions are less likely to be triggered. This effect additionally suppresses leakage.

Having validated the truncated Hamiltonian for our analysis, we re-express Eq~(\ref{eq:trunc_H_spectrometer}) in terms of the Pauli operators for simplicity
\begin{align}
    \label{Supp_eq:H_SL_eff}
    \tilde{H}_{\mathrm{SL}}^{(j-1,j)} = \frac{\hbar}{2}\Omega^{(j-1,j)} \tilde{\sigma}_z^{(j-1,j)} +\hbar \left[ \tilde{B}^{(j-1,j)}_{\perp}(t) (\tilde{\sigma}_{+}^{(j-1,j)} +\tilde{\sigma}_{-}^{(j-1,j)}) + \tilde{B}^{(j-1,j)}_{\parallel}(t)\left(\frac{\tilde{\sigma}_z^{(j-1,j)}}{2}\right) \right],
\end{align}
where $\tilde{\sigma}_z^{(j-1,j)}$, $\tilde{\sigma}_+^{(j-1,j)}$, and $\tilde{\sigma}_-^{(j-1,j)}$ denote the Pauli Z, raising, and lowering operators of the $j$-th spectrometer, respectively. We define the noise operators $\tilde{B}^{(j-1,j)}_{\perp}(t)$, $\tilde{B}^{(j-1,j)}_{\parallel}(t)$, describing longitudinal and transverse relaxation in the $j$-th spin locking subspace, respectively. They are given as linear combinations of the noise operators $B^{(j)}(t)$
\begin{align}
    \label{Supp_eq:B_parallel_general}
    \tilde{B}^{(j-1,j)}_{\perp}(t) = \sum_{k=1}^{d-1} B^{(k)}(t) \langle j-1 | V^{\dagger} |k \rangle \langle k| V |j\rangle  = \sum_{k=1}^{d-1} B^{(k)}(t) \langle j | V^{\dagger} |k \rangle \langle k| V |j-1\rangle
    \equiv \sum_{k=1}^{d-1} \alpha^{(k)}_{(j-1,j)} B^{(k)}(t), 
\end{align}
\begin{align}
    \label{Supp_eq:B_longitudinal_general}
    \tilde{B}^{(j-1,j)}_{\parallel}(t) = \sum_{k=1}^{d-1} B^{(k)}(t) \Big[\langle j-1 | V^{\dagger} |k \rangle \langle k| V |j-1\rangle - \langle j | V^{\dagger} |k \rangle \langle k| V |j\rangle \Big] \equiv  \sum_{k=1}^{d-1} \beta^{(k)}_{(j-1,j)} B^{(k)}(t),
\end{align} 
where the noise participation ratio  $\alpha^{(k)}_{(j-1,j)}$ ($\beta_{(j-1,j)}^{(k)}$) is a dimensionless factor that quantifies the fraction of the energy fluctuation for level $k$ that translates to the transverse (longitudinal) noise. The change-of-basis matrix $V$ contains information about how the bare states mix and form the dressed states. By numerically estimating the matrix $V$, we calculated $\alpha^{(k)}_{(j-1,j)}$ as a function of $A_{\mathrm{drive}}$ (Fig.~3).

\section{Derivation of Reduced Master Equation}
\label{SuppSec:ReducedMasterEq}

We follow Ref.~[S8] to derive a reduced master equation for the spin locking states of the $j$-th spectrometer. Note that the standard spin-locking theory discussed in Ref.~[S8] does not include the longitudinal noise in the spin locking frame, from the multi-level nature of the sensor, as detailed above. 

The starting point of this derivation is the Hamiltonian of the $j$-th spectrometer (Eq.~\eqref{Supp_eq:H_SL_eff}).
\begin{align}
    {H}_{\mathrm{SL}} = \frac{\hbar}{2}\Omega {\sigma}_z +\hbar \left[ {B}_{\perp}(t) ({\sigma}_{+} +{\sigma}_{-}) + {B}_{\parallel}(t)\left(\frac{{\sigma}_z}{2}\right) \right] \equiv H_{\mathrm{S}}' + H_{\mathrm{SB}}',
\end{align}
Here, we have omitted the superscripts $(j-1,j)$ and the tildes for simplicity, and introduced $H_{\mathrm{S}}' \equiv \frac{\hbar}{2}\Omega {\sigma}_z$ and $H_{\mathrm{SB}}' \equiv \hbar \left[ {B}_{\perp} ({\sigma}_{+} + \sigma_{-}) + B_{\parallel}\left(\sigma_z/2\right) \right]$, which denote the spectrometer Hamiltonian and the spectrometer-bath interaction Hamiltonian respectively. 

To employ the time convolutionless (TCL) projection operator technique~[S8-S9], we move to the interaction picture with respect to the system Hamiltonian $H_{\mathrm{S}}'$. Then, we have the interaction Hamiltonian ${H}_{\mathrm{SB}}$ as follows:
\begin{align}
    \label{supp_eq:H_int}
    {H}_{\mathrm{SB}}(t) = \hbar \left[ {B}_{\perp}(t) \left[e^{i\Omega t}\sigma_+ + e^{-i\Omega t}\sigma_-\right] + \frac{1}{2} {B}_{\parallel} (t)\sigma_z \right].
\end{align}
To employ the technique, we assume the dephasing noise ${B}^{(j)}(t)$ is stationary noise with zero mean such that $\langle{B}^{(j)}(t) \rangle = 0$, $\langle{B}^{(j)}(t_1)\tilde{B}^{(j)}(t_2) \rangle =  \langle {B}^{(j)}(t_1-t_2)\tilde{B}^{(j)}(0) \rangle$ and the coupling between the system and the bath is weak enough to truncate the time convolutionless generator at second order.

In this interaction picture, the equation of motion for the density matrix of the total system ${\rho}_{\mathrm{tot}}(t)$ is given as follows:
\begin{align}
    \frac{\partial}{\partial t} {\rho}_{\mathrm{tot}}(t) = -\frac{i}{\hbar} [H_{\mathrm{SB}}(t), {\rho}_{\mathrm{tot}}(t)] \equiv \mathcal{L}(t){\rho}_{\mathrm{tot}}(t).
\end{align}
The Liouville super-opeartor $\mathcal{L}(t)$ has been defined as $\mathcal{L}(t)\cdot\equiv -i/\hbar [ H_{\mathrm{SB}}(t), \cdot]$.

To apply the TCL technique, we assume the initial state of the total system is separable such that ${\rho}_{\mathrm{tot}}(0) = {\rho}_{\mathrm{S}}(0)\otimes \rho_{B}(0)$, where $\rho_{S}(t)$ and $\rho_{B}(t)$ denote the density matrix of the spectrometer and the bath in the interaction picture at time $t$, respectively. Now we introduce the projection superoperator 
\begin{align}
 \mathcal{P}\cdot \equiv \rho_{\mathrm{B}}(0) \mathrm{Tr}_{\mathrm{B}} [\cdot],
\end{align}
which projects on the sensor part of the density matrix ${\rho}_{\mathrm{tot}}$, where $\mathrm{Tr}_{\mathrm{B}}$ denotes the tracing out operation over the bath. A complementary projection superoperator $\mathcal{Q}$ is then defined by $\mathcal{Q} \equiv I -\mathcal{P}$, where $I$ is the identity superoperator. Using the above projection operators, we can write down the TCL master equation truncated at second order as follows:
\begin{align}
    \label{supp_eq:Prho_tot}
    \frac{\partial}{\partial t} \mathcal{P}{\rho}_{\mathrm{tot}}(t) = \mathcal{K}(t) \mathcal{P}{\rho}_{\mathrm{tot}}(t),
\end{align}
where $\mathcal{K}(t)$ denotes the second-order TCL generator and is given as
\begin{align}
    \mathcal{K}(t) = \int_{0}^{t}\diff{s} \mathcal{P}\mathcal{L}(t)\mathcal{Q}\mathcal{L}(s) {\rho}_{\mathrm{tot}}(t).
\end{align}

For noise with vanishing mean ($\langle \tilde{B}_{\perp} (t) \rangle = \langle \tilde{B}_{\parallel} (t) \rangle = 0$), substituting Eq.~(\ref{supp_eq:H_int})
into Eq.~(\ref{supp_eq:Prho_tot}) and tracing over the bath leads to the following equation
\begin{align}
    \frac{\partial}{\partial t}{\rho}(t) = \int_{0}^{t}\diff{s} \Bigg[& C_{\perp}(t,s) \left[e^{i\Omega (s-t)} \left[  \sigma_{+} {\rho}(t) \sigma_{-} - \sigma_{-}\ \sigma_{+}{\rho}(t) \right] +  e^{i\Omega (t-s)} \left[  \sigma_{-} {\rho}(t) \sigma_{+} - \sigma_{+}\ \sigma_{-}{\rho}(t)\right] \right] 
    \nonumber \\
    &+ C_{\perp}(s,t) \left[e^{i\Omega (s-t)} \left[  \sigma_{-} {\rho}(t) \sigma_{+} - {\rho}(t)\sigma_{+}\ \sigma_{-} \right] +  e^{i\Omega (t-s)} \left[  \sigma_{+} {\rho}(t) \sigma_{-} - {\rho}(t) \sigma_{-} \sigma_{+}\right] \right]
    \nonumber \\
    &+ \frac{1}{4} C_{\parallel}(t,s) \left[ \sigma_z {\rho}(t) \sigma_z - \sigma_z \sigma_z {\rho}(t) \right] 
    + \frac{1}{4} C_{\parallel}(s,t) \left[ \sigma_z {\rho}(t) \sigma_z - {\rho}(t) \sigma_z \sigma_z \right] \Bigg],
\end{align}
which describes the evolution of the reduced density matrix of the system, ${\rho}(t) = \mathrm{Tr}_{\mathrm{B}} {\rho}_{\mathrm{tot}}(t)$, in terms of the correlation functions
\begin{align}
C_{\perp} (t,s) = \mathrm{Tr}_{\mathrm{B}}\left[  {B}_{\perp} (t) {B}_{\perp} (s)  \rho_{\mathrm{B}}\right],
\hspace{5mm}
C_{\parallel} (t,s) = \mathrm{Tr}_{\mathrm{B}}\left[  {B}_{\parallel} (t) {B}_{\parallel} (s)  \rho_{\mathrm{B}}\right].
\end{align}
Since we assumed that the bath-induced noise is a stationary process ($C_{\perp}(t,s) = C_{\perp}(t-s)$, $C_{\parallel}(t,s) = C_{\parallel}(t-s)$), we can substitute $s\equiv t-t'$, and rewrite the equation in the frequency domain as follows:
\begin{align}
    \frac{\partial}{\partial t}{\rho}(t) = \int_{0}^{t}\diff{t'}\frac{1}{2\pi}\int_{-\infty}^{\infty}\diff{\omega} e^{i\omega t'} \Bigg[ &
    S_{\perp}(\omega)\left[ e^{-i\Omega t'}\left[ \sigma_{+} {\rho}(t)\sigma_{-} -\sigma_{-}\sigma_{+}{\rho}(t) \right] +e^{i\Omega t'}\left[ \sigma_{-} {\rho}(t)\sigma_{+} -\sigma_{+}\sigma_{-}{\rho}(t) \right] \right]
    \nonumber \\
    &+S_{\perp}(-\omega)\left[ e^{-i\Omega t'}\left[ \sigma_{-} {\rho}(t)\sigma_{+} -{\rho}(t)\sigma_{+}\sigma_{-} \right] +e^{i\Omega t'}\left[ \sigma_{+} {\rho}(t)\sigma_{-} -{\rho}(t)\sigma_{-}\sigma_{+} \right] \right]
    \nonumber \\
    &+\frac{1}{4}S_{\parallel}(\omega) \left[ \sigma_z {\rho}(t) \sigma_z - \sigma_z \sigma_z {\rho}(t) \right] 
    +\frac{1}{4}S_{\parallel}(-\omega) \left[ \sigma_z {\rho}(t) \sigma_z - {\rho}(t) \sigma_z \sigma_z  \right] 
    \Bigg],
\end{align}
where we have introduced the power spectral densities of noise operators, respectively as follows:
\begin{align} 
    S_{\perp}(\Omega) &= \int_{-\infty}^{\infty}\diff{\tau} e^{-i\Omega \tau} \langle {B}_{\perp}(\tau){B}_{\perp}(0)\rangle, \\
    S_{\parallel}(\Omega) &= \int_{-\infty}^{\infty}\diff{\tau} e^{-i\Omega \tau} \langle {B}_{\parallel}(\tau){B}_{\parallel}(0)\rangle.
\end{align}
Following Ref.~[S9], we rewrite the equation by introducing a filter function $F(\omega)\equiv \int_{0}^{t}e^{i\omega t'}\diff{t'}$ for the free induction decay~[S11]. 

\begin{align}
\label{supp_eq:master_eq_filter_func}
    \frac{\partial}{\partial t}{\rho}(t) = \frac{1}{2\pi} \int_{-\infty}^{\infty}\diff{\omega} \Bigg[ &
    S_{\perp}(\omega)\Big[ F(\omega-\Omega)\left[ \sigma_{+} {\rho}(t)\sigma_{-} -\sigma_{-}\sigma_{+}{\rho}(t) \right] +F(\omega+\Omega)\left[ \sigma_{-} {\rho}(t)\sigma_{+} -\sigma_{+}\sigma_{-}{\rho}(t) \right] \Big]
    \nonumber \\
    &+S_{\perp}(-\omega)\Big[ F(\omega-\Omega)\left[ \sigma_{-} {\rho}(t)\sigma_{+} -{\rho}(t)\sigma_{+}\sigma_{-} \right] +F(\omega+\Omega)\left[ \sigma_{+} {\rho}(t)\sigma_{-} -{\rho}(t)\sigma_{-}\sigma_{+} \right] \Big]
    \nonumber \\
    &+\frac{1}{4}S_{\parallel}(\omega) F(\omega)\left[ \sigma_z {\rho}(t) \sigma_z - \sigma_z \sigma_z {\rho}(t) \right] 
    +\frac{1}{4}S_{\parallel}(-\omega) F(\omega)\left[ \sigma_z {\rho}(t) \sigma_z - {\rho}(t) \sigma_z \sigma_z  \right] 
    \Bigg]
\end{align}
The filter function $F(\omega)$ acts as a bandpass filter for the noise spectra, which is peaked at $\omega = 0$ with $1/t$ bandwidth. Assuming that all spectra vary negligibly over this passband, we approximate $F(\omega)$ by the delta function as follows:
\begin{align}
    \lim_{t\rightarrow \infty} F(\omega) = \pi \delta (\omega).
\end{align}
Then, the equation~(\ref{supp_eq:master_eq_filter_func}) can be written as:
\begin{align}
    \frac{\partial}{\partial t}{\rho}(t) = &
    S_{\perp} (\Omega)\Big[\sigma_{+} {\rho}(t)\sigma_{-}  - \frac{1}{2}\left[ \sigma_{-} \sigma_{+}{\rho}(t) +{\rho}(t)\sigma_{-}\sigma_{+} \right] \Big]
    \nonumber \\
    &+S_{\perp} (-\Omega)\Big[  \sigma_{-} {\rho}(t)\sigma_{+}   -\frac{1}{2}\left[ \sigma_{+}\sigma_{-} {\rho}(t) +{\rho}(t)\sigma_{+}\sigma_{-} \right] \Big]
    \nonumber \\
    &+\frac{1}{4}S_{\parallel} (0)\left[\sigma_z {\rho}(t) \sigma_z - \frac{1}{2}\left[\sigma_z \sigma_z {\rho}(t) + {\rho}(t)\sigma_z \sigma_z \right] \right]
\end{align}
From this equation, we find that the longitudinal decay rate $\Gamma_{1\rho}^{(j-1,j)}$ and the equilibrium-state polarization $\langle \tilde{\sigma}_z(t)  \rangle|_{t\rightarrow\infty}$ in the spin-locking frame will be given as follows (we put the superscripts $(j-1,j)$ and tildes back for the sake of consistency with the main text):
\begin{align}
\label{Supp_eq:Gamma_1rho_final}
\Gamma_{1\rho}^{(j-1,j)}  = \tilde{S}_{\perp} (\Omega^{(j-1,j)})+\tilde{S}_{\perp} (-\Omega^{(j-1,j)}) \\ 
\label{Supp_eq:sigma_z_final}
\langle \tilde{\sigma}_z(t)  \rangle|_{t\rightarrow\infty} = \frac{\tilde{S}_{\perp} (\Omega^{(j-1,j)})-\tilde{S}_{\perp} (-\Omega^{(j-1,j)})}{\tilde{S}_{\perp} (\Omega^{(j-1,j)}) + \tilde{S}_{\perp} (-\Omega^{(j-1,j)})}
\end{align}

This result shows that transverse relaxation rate in the multi-level spin-locking experiment can be used to determine the noise spectral density of the longitudinal noise in the laboratory frame in the same way as for an ideal two-level system. Remarkably, the longitudinal noise PSD $\tilde{S}_{\parallel}$, which is an artifact of the multi energy level structure, does not enter Eqs.~(\ref{Supp_eq:Gamma_1rho_final}, \ref{Supp_eq:sigma_z_final}), which are consistent with the standard spin-locking analysis.

{
\section{
Contributions of $T_1$ decay to the spin relaxation  $\Gamma_{1\rho}$}
\label{SuppSec:T1contribution}
As discussed in Ref.~[S12], energy relaxation ($T_1$ decay) of a two-level qubit sensor contributes to the longitudinal relaxation of the spin, locked at Rabi frequency $\Omega$. This $T_1$ contribution to the spin relaxation can be expressed by the following equation:
\begin{align}
    \label{eq:T1_contribution_qubit}
    \Gamma_{1\rho}(\Omega)=\frac{1}{2}\Gamma_1+\Gamma_{\varphi}(\Omega) = \frac{1}{2}(\Gamma_{1\uparrow}+\Gamma_{1\downarrow}) + \Gamma_{\varphi}(\Omega).
\end{align}
Here $\Gamma_{1\rho}(\Omega)$ denotes the longitudinal spin relaxation rate, and $\Gamma_{\varphi}(\Omega)$ denotes the relaxation rate due to pure dephasing noise PSD at the locking Rabi frequency $\Omega$. The energy relaxation rate $\Gamma_1 = 1/T_1$ of the qubit is given as a sum of an ``up transition rate'' $\Gamma_{1\uparrow}$ (from $|0\rangle$ to $|1\rangle$), and a ``down transition rate'' $\Gamma_{1\downarrow}$ ($|1\rangle$ to $|0\rangle$)~[S13]. These up and down transition rates $\Gamma_{1\uparrow}$, $\Gamma_{1\downarrow}$ are determined by the noise spectrum $S_x(\omega)$ causing the energy relaxation at the qubit frequency $\omega_q$~[S8], 
\begin{align}
    \Gamma_{1\uparrow} = S_x(-\omega_q) \text{ and }  \Gamma_{1\downarrow} = S_x(+\omega_q).
\end{align}
Typically, qubits operate at low temperature and are assumed to be in thermal equilibrium with its cryogenic environment ($k_B T \ll \hbar \omega_q$). In this case, according to Maxwell-Boltzmann statistics, the up-rate $\Gamma_{1\uparrow}$ is exponentially smaller than the down-rate $\Gamma_{1\downarrow}$ by the Boltzmann factor $\Gamma_{1\uparrow}/\Gamma_{1\downarrow} = \exp{\left[-(\hbar \omega_q) /(k_B T) \right]}$. Therefore, in the low-temperature limit, we can approximate the relaxation rate $\Gamma_1$ as
\begin{align}
    \label{eq:T1_low_temp}
    \Gamma_1 \approx \Gamma_{1\downarrow} = S_x(+\omega_q).
\end{align}
Note that Eq.~(\ref{eq:T1_contribution_qubit}) relies on the assumption that the noise spectrum at qubit frequency $S_x(\pm\omega_q)$ varies negligibly within $\pm$ Rabi frequency $\pm\Omega$, such that $  S_x(\omega_q) \simeq S_x(\omega_q \pm \Omega)$, and $S_x(-\omega_q) \simeq S_x(-\omega_q \pm \Omega)$~[S8].

Similarly, in the case of a multi-level sensor, we can apply Eq.~(\ref{eq:T1_contribution_qubit}) to the $j$-th spin-locked sensor in the weak driving limit (small $\Omega^{(j-1,j)}$),
\begin{align}
    \Gamma_{1\rho}^{(j-1,j)}(\Omega^{(j-1,j)}) = \frac{1}{2}\Gamma_{1}^{(j-1,j)} +\Gamma_{\varphi}(\Omega^{(j-1,j)}) = \frac{1}{2}\left( \Gamma_{1\uparrow}^{(j-1,j)}  + \Gamma_{1\downarrow}^{(j-1,j)} \right) +  \left(\tilde{S}_{\perp}(\Omega^{(j-1,j)}) + \tilde{S}_{\perp}(-\Omega^{(j-1,j)})\right) \nonumber \\ ,
\end{align}
where $\Gamma_{1}^{(j-1,j)}\equiv1/T_1^{(j-1,j)}$ denotes the the energy relaxation rate between $|j-1\rangle$ and $|j\rangle$. Note that the above equation is valid only if the locking Rabi frequency, $\Omega^{(j-1,j)}$, is much smaller than the level anharmonicity of the sensor. 
If the $\Omega^{(j-1,j)}$ is comparable to level anharmonicity, then multi-level dressing effect must be taken into account.

To consider the multi-level dressing effect, we follow a similar approach to what we discussed in Supplementary Material 4. We estimate the matrix elements of the $k$-th raising and lowering operator, $\sigma_{+}^{(k-1,k)}\equiv|k\rangle \langle k-1|, \sigma_{-}^{(k-1,k)}\equiv|k-1\rangle \langle k|$, which are associated with the energy relaxation between $|k\rangle$ and $|k-1\rangle$, for the transition between the $j$-th spin-locked states: $\langle+^{(j-1,j)}| \sigma_{\pm}^{(k-1,k)}|-^{(j-1,j)}\rangle$ and $\langle-^{(j-1,j)}| \sigma_{\pm}^{(k-1,k)}|+^{(j-1,j)}\rangle$. Notably, these matrix elements can be estimated by using the change-of-basis matrix $V$, introduced in Supplementary Material 4. For instance, $\langle+^{(j-1,j)}| \sigma_{+}^{(k-1,k)}|-^{(j-1,j)}\rangle$ can be estimated as follows:
\begin{align}
    \langle+^{(j-1,j)}| \sigma_{+}^{(k-1,k)}|-^{(j-1,j)}\rangle = \langle j| V^{\dagger}  \left(|k\rangle \langle k-1| \right) V |j-1\rangle.
\end{align}
Remarkably, the matrix elements for the transition between the $j$-th spin-locked states tell us at what rate the raising and lowering operator $\sigma_{\pm}^{(k-1,k)}$ (which are associated with $T_1$) will cause transitions between the spin-locked states. Namely, we can estimate the effective longitudinal spin relaxation rate, $\Gamma_{1,\mathrm{eff}}^{(j-1,j)}$, as a consequence of the energy relaxation of a multi-level qubit, as follows:
\begin{align}
    \label{eq:Gamma_1_eff}
    \Gamma_{1,\mathrm{eff}}^{(j-1,j)} = 
  \sum_{k} &\langle|j-1| V^{\dagger}  \left( \Gamma_{1\downarrow}^{(k-1,k)} |k-1\rangle \langle k| +\Gamma_{1\uparrow}^{(k-1,k)} |k\rangle \langle k-1| \right) V |j\rangle \nonumber \\ 
     & + \langle|j| V^{\dagger}  \left( \Gamma_{1\downarrow}^{(k-1,k)} |k-1\rangle \langle k| + \Gamma_{1\uparrow}^{(k-1,k)} |k\rangle \langle k-1| \right) V |j-1\rangle \nonumber \\
   =\sum_{k} &\langle|j-1| V^{\dagger}  \left( S_x(+\omega_s^{(k-1,k)}) |k-1\rangle \langle k| + S_x(-\omega_s^{(k-1,k)}) |k\rangle \langle k-1| \right) V |j\rangle \nonumber \\ 
   & + \langle|j| V^{\dagger}  \left( S_x(+\omega_s^{(k-1,k)}) |k-1\rangle \langle k| + S_x(-\omega_s^{(k-1,k)}) |k\rangle \langle k-1| \right) V |j-1\rangle.
\end{align}
Assuming the low effective temperature for the multi-level sensor ($\hbar\omega_{s}^{(k-1,k)}\gg k_B T $, we can replace $\Gamma_{1\downarrow}^{(k-1,k)} = S_x(+\omega_s^{(k-1,k)})\approx \Gamma_1^{(k-1,k)}$, and $\Gamma_{1\uparrow}^{(k-1,k)} = S_x(-\omega_s^{(k-1,k)})\approx 0$. Under this approximation Eq.~(\ref{eq:Gamma_1_eff}) can be rewritten as:

\begin{align}
    \label{eq:Gamma_1_eff}
    \Gamma_{1,\mathrm{eff}}^{(j-1,j)} \approx \sum_{k} \Gamma_{1}^{(k-1,k)} \left( \langle j-1| V^{\dagger}  \left(|k-1\rangle \langle k| \right) V |j\rangle \nonumber + \langle j| V^{\dagger}  \left(|k-1\rangle \langle k| \right) V |j-1\rangle \right).
\end{align}
Finally, the overall longitudinal relaxation rate $\Gamma_{1\rho}(\Omega)$ for a multi-level sensor, including the $T_1$ decay contribution, is given as
\begin{align}
     \Gamma_{1\rho}^{(j-1,j)}(\Omega)=\frac{1}{2}\Gamma_{1,\mathrm{eff}}^{(j-1,j)}+\Gamma_{\varphi}(\Omega).
\end{align}
Note that if the energy relaxation rate is much faster than the dephasing rate at Rabi frequency $\Omega^{(j-1,j)}$ ($\Gamma_{1,\mathrm{eff}}^{(j-1,j)} \gg \Gamma_{\varphi}(\Omega)$), the $T_1$ contribution can overshadow the contribution of dephasing noise. This $T_1$ overshadowing effect can make it hard to extract the dephasing noise, especially if $T_1$ is fluctuating. Namely, $T_1$ of the qubit sensor limits the noise sensitivity (= the smallest dephasing noise that can be detected reliably).  

}

\section{Separating out the contributions of $T_1$ decay and native dephasing}
\label{SuppSec:IsolatingT1}
In the $\mathrm{SL}^{(0,1)}$ and $\mathrm{SL}^{(1,2)}$ experiments, we measure the longitudinal spin relaxation in both the presence and absence of the engineered noise to separate out the contributions of $T_1$ decay of the transmon sensor and native dephasing ($T_{\varphi}$) from the estimation of $\tilde{S}^{(j-1,j)}_{\perp}(\omega)$. The longitudinal spin relaxation rate in the presence of engineered noise for the $j$-th spin-locked spectrometer 
, $\Gamma_{1\rho}^{(j-1,j), \mathrm{pres}}$ is given as
\begin{align}
\label{SuppEq:Gamma1rho_w}
\Gamma_{1\rho}^{(j-1,j), \mathrm{pres}} = \left( \tilde{S}_{\perp}^{(j-1,j)} (\Omega^{(j-1,j)})+\tilde{S}_{\perp} (-\Omega^{(j-1,j)}) \right)+ \left( \tilde{S}_{\perp}^{\varphi}(\Omega^{(j-1,j)})  +  \tilde{S}_{\perp}^{\varphi} (-\Omega^{(j-1,j)}) \right) + \Gamma_1^{(j-1,j)}/2,
\end{align} 
where $\tilde{S}_{\perp}^{\varphi} (\Omega)$ corresponds to the power spectral density of the longitudinal spin relaxation noise, which is contributed from the native dephasing noise. The contribution of $T_1$ decay to the spin relaxation is denoted by $(\Gamma_1/2)$. Note that, for small $\Omega$ compared to the anharmonicity of the sensor, $\Gamma_1^{(j-1,j)} $ can be approximated as $1/T_1^{(j-1,j)}$, where $T_1^{(j-1,j)}$ is the lab-frame longitudinal relaxation time of the sensor for $|j-1\rangle$--$|j\rangle$ transition. As $\Omega$ increases, the peripheral states (neither $|j-1\rangle$ nor $|j\rangle$) start to participate to form the spin-locked states $\{ |+\rangle^{(j-1,j)}, |-\rangle^{(j-1,j)} \}$. Hence, the longitudinal relaxation for the peripheral level transitions also start to contribute to $\Gamma_1^{(j-1,j)}$ as $\Omega^{(j-1,j)}$ increases. This contribution can be numerically accounted for by considering multi-level dressing as discussed in Sec. \ref{SuppSec:Derivation_Heff}. 

The spin-polarization at equilibrium in the presence of engineered noise $\langle \sigma_{z}^{(j-1,j)}(t) \rangle|^{\mathrm{pres}}_{t\rightarrow\infty}$ is given by
\begin{align}
\langle \tilde{\sigma}_{z}^{(j-1,j)}(t) \rangle|^{\mathrm{pres}}_{t\rightarrow\infty} = \frac{\tilde{S}_{\perp} (\Omega^{(j-1,j)})-\tilde{S}_{\perp} (-\Omega^{(j-1,j)}) + \tilde{S}_{\perp}^{\varphi} (\Omega^{(j-1,j)})  -\tilde{S}_{\perp}^{\varphi} (-\Omega^{(j-1,j)}) }{\tilde{S}_{\perp} (\Omega^{(j-1,j)}) + \tilde{S}_{\perp} (-\Omega^{(j-1,j)}) +\tilde{S}_{\perp}^{\varphi} (\Omega^{(j-1,j)}) +  \tilde{S}_{\perp}^{\varphi} (-\Omega^{(j-1,j)})}.
\end{align}
Accordingly, in the absence of the engineered noise, the spin-locking relaxation rate ($\Gamma_{1\rho}^{(j-1,j), \mathrm{abs}}$) and the equilibrium spin-polarization ($\langle \tilde{\sigma}_z^{(j-1,j)}(t) \rangle|^{\mathrm{abs}}_{t\rightarrow\infty}$) are given as follows:
\begin{align}
\label{SuppEq:Gamma1rho_wo}
\Gamma_{1\rho}^{(j-1,j), \mathrm{abs}} = \tilde{S}_{\perp}^{\varphi} (\Omega^{(j-1,j)})+\tilde{S}_{\perp}^{\varphi} (-\Omega^{(j-1,j)}) + \Gamma^{(j-1,j)}_1/2,
\end{align} 
\begin{align}
\langle \tilde{\sigma}^{(j-1,j)}_z(t) \rangle|^{\mathrm{abs}}_{t\rightarrow\infty} = \frac{\tilde{S}_{\perp}^{\varphi} (\Omega^{(j-1,j)})-\tilde{S}_{\perp}^{\varphi} (-\Omega^{(j-1,j)}) }{\tilde{S}_{\perp}^{\varphi} (\Omega^{(j-1,j)}) + \tilde{S}_{\perp}^{\varphi} (-\Omega^{(j-1,j)})}.
\end{align}
Assuming that the engineered dephasing noise is much stronger than the native dephasing noise $\big(\tilde{S}_{\perp} (\Omega^{(j-1,j)}) + \tilde{S}_{\perp} (-\Omega^{(j-1,j)})\gg \tilde{S}_{\perp}^{\varphi} (\Omega) +  \tilde{S}_{\perp}^{\varphi} (-\Omega^{(j-1,j)})\big)$, the $\langle \tilde{\sigma}_{z}^{(j-1,j)}(t) \rangle|^{\mathrm{pres}}_{t\rightarrow\infty} $ can be approximated as follows:
\begin{align}
\label{SuppEq:sz_w_approx}
\langle \tilde{\sigma}^{(j-1,j)}_{z}(t) \rangle|^{\mathrm{pres}}_{t\rightarrow\infty} &\approx  \frac{\tilde{S}_{\perp} (\Omega^{(j-1,j)})-\tilde{S}_{\perp} (-\Omega^{(j-1,j)}) + \tilde{S}_{\perp}^{\varphi} (\Omega^{(j-1,j)})  -\tilde{S}_{\perp}^{\varphi} (-\Omega^{(j-1,j)}) }{\tilde{S}_{\perp} (\Omega^{(j-1,j)}) + \tilde{S}_{\perp} (-\Omega^{(j-1,j)})} \nonumber \\
&\approx \frac{\tilde{S}_{\perp} (\Omega^{(j-1,j)})-\tilde{S}_{\perp} (-\Omega^{(j-1,j)})}{\tilde{S}_{\perp} (\Omega^{(j-1,j)}) + \tilde{S}_{\perp} (-\Omega^{(j-1,j)})} +\frac{\tilde{S}^{\varphi}_{\perp} (\Omega^{(j-1,j)})+\tilde{S}^{\varphi}_{\perp} (-\Omega^{(j-1,j)})}{\tilde{S}_{\perp} (\Omega^{(j-1,j)}) + \tilde{S}_{\perp} (-\Omega^{(j-1,j)})} \langle \tilde{\sigma}_z^{(j-1,j)}(t) \rangle|^{\mathrm{abs}}_{t\rightarrow\infty}
\nonumber \\
&\approx \frac{\tilde{S}_{\perp} (\Omega^{(j-1,j)})-\tilde{S}_{\perp} (-\Omega^{(j-1,j)})} {\tilde{S}_{\perp} (\Omega^{(j-1,j)})+\tilde{S}_{\perp} (-\Omega^{(j-1,j)})}
\end{align}
From Eqs.~(\ref{SuppEq:Gamma1rho_w}, \ref{SuppEq:Gamma1rho_wo}, and \ref{SuppEq:sz_w_approx}), we can solve for the  $\tilde{S}_{\perp}(\Omega^{(j-1,j)})$ as follows: 
\begin{align}
    \tilde{S}_{\perp}(\Omega^{(j-1,j)}) = \frac{(1+\langle \tilde{\sigma}_{z}^{(j-1,j)}(t) \rangle|^{\mathrm{pres}}_{t\rightarrow\infty} )}{2} \left(\Gamma_{1\rho}^{(j-1,j), \mathrm{pres}}-\Gamma_{1\rho}^{(j-1,j), \mathrm{abs}}\right)
\end{align}

In addition, $T_1$ decay of the transmon sensor (mostly for $|j-2\rangle$--$|j-1\rangle$ transition) also results in the leakage of the $j$-th spin-locked states from the $j$-th spin-locking subspace. We effectively separated out this $T_1$-induced leakage from the the estimation of $\tilde{S}_{\perp}(\Omega^{(j-1,j)})$ by measuring the normalized spin polarization $\langle \tilde{\sigma}^{(j-1,j)}_z (t) \rangle =  \frac{\rho^{(j-1,j-1)}(t) - \rho^{(j,j)}(t) }{\rho^{(j-1,j-1)}(t) + \rho^{(j,j)}(t)}$. Namely, we  compensated the spin polarization for leakage of the spin-locked states $|+\rangle^{(j-1,j)}$ and $|-\rangle^{(j-1,j)}$ by multiplying a factor of $1/\big(\rho^{(j-1,j-1)}(\tau) + \rho^{(j,j)}(\tau)\big)$ to the spin-polarization, $\left(\rho^{(j-1,j-1)}(t) - \rho^{(j,j)}(t)\right)$. 

\section{Advantages of the spin-locking QNS over the dynamic decoupling QNS}
Dynamic decoupling (D.D.) noise spectroscopy during predominantly free evolution has been widely used to characterize dephasing noise in various qubit systems. However, even within a two-level approximation, there are advantages to using a spin-locking (driven evolution) approach to noise spectroscopy. We discuss these advantages here to justify why we focus on extending the spin-locking based multi-level QNS throughout this work.

First, dynamic decoupling noise spectroscopy uses multiple, ideally instantaneous control pulses, which are used to flip the qubit states in the time domain and thereby realize a desired filter function in the frequency domain. Although we often think of these as narrow filters---ideally delta functions---that can sample the noise at any particular frequency, in practice, the width of the filter is determined by both the number of pulses and the duration of the experiment. Narrow filters require large numbers of pulses, which are not instantaneous in practice and take up available free-evolution time, and therefore lead to longer experiments (fighting against $T_1$). In turn, high-frequency spectroscopy requires small time separation between pulses, yet high-fidelity pulses are not boxcars, but generally have Gaussian or cosine envelopes, requiring a minimum time between pulses to remain accurate.  Taken together, this trade space can be somewhat contradictory and may make high-frequency noise spectroscopy with high precision a challenge. Spin locking does not have these issues, because it is essentially a single quasi-continuous drive. 

In practice, the D.D. control pulses are generally imperfect [S14, S15]. These imperfections in control pulses (= control errors) limit the spectral range of the sensor, as discussed in Ref. [S16]. Namely, the spectral range is mainly limited by control imperfections, rather than physical constraints inherent in the system. In contrast, the spin-locking noise spectroscopy does not require fast-pulsed control (it is a single quasi-CW drive); therefore, it can characterize higher frequency noise than the D.D. based noise spectroscopy as shown in Ref. [S12] (this is also related to first point above). 

In addition, practical D.D. filters have a rather broad bandwidth (compared with spin locking), and a deconvolution step is required to extract the noise about some frequency.  Furthermore, in many common approaches like the CPMG sequence, the filter function has one predominant peak and smaller lobes away from this peak, which exacerbates the deconvolution problem (a notable exception is the use of Slepian pulses, but these still lead to relatively broad filters). In contrast, the spin-locking approach uses a single, continuous drive, and we simply monitor decoherence times within the driven qubit basis to extract noise at the Rabi frequency with relatively small “filter bandwidth” (limited by coherence). This approach is much simpler to implement and generally more precise.

\subsection*{Supplementary References}

\begin{itemize}

\item[{[S1]}] R. Barends, J. Kelly, A. Megrant, D. Sank, E. Jeffrey, Y. Chen, Y. Yin, B. Chiaro, J.  Mutus, C. Neill, P. O’Malley, P. Roushan, J. Wenner, T. C. White, A. N. Cleland, and  J. M. Martinis. Coherent Josephson qubit Suitable for scalable quantum integrated circuits. \textit{Phys. Rev. Lett.} \textbf{111}, 080502 (2013).

\item[{[S2]}] J. Koch, T. M. Yu, J. Gambetta,  A.  A. Houck, D. I. Schuster, J. Majer, A. Blais, M. H. Devoret, S. M. Girvin, and R. J. Schoelkopf. Charge-insensitive qubit design derived from the Cooper pair box. \textit{Phys. Rev. A} \textbf{76}, 042319 (2007).

\item[{[S3]}]  E. Jeffrey, D. Sank, J. Y. Mutus, T. C. White, J. Kelly, R. Barends, Y. Chen, Z. Chen, B. Chiaro, A. Dunsworth, A.  Megrant,  P. J.  O’Malley, C. Neill, P. Roushan, A. Vainsencher, J.  Wenner, A. N. Cleland,  and J. M. Martinis. Fast accurate state measurement with superconducting qubits. \textit{Phys. Rev. Lett.} \textbf{112}, 190504 (2014).

\item[{[S4]}] E. A. Sete, J. M. Martinis, and A. N. Korotkov. Quantum theory of a bandpass Purcell filter for qubit readout. \textit{Phys. Rev. A} \textbf{92}, 012325 (2015).

\item[{[S5]}] Y. Sung, F. Beaudoin, L. M. Norris, F. Yan, D. Kim, J. Y. Qiu, U.V. L{\"{u}}pke, J. L. Yoder, T. P. Orlando, S. Gustavsson, L. Viola, and  W. D. Oliver. Non-Gaussian noise spectroscopy with a superconducting qubit sensor. \textit{Nature Communications} \textbf{10}, 3715 (2019).

\item[{[S6]}] C. Macklin, K. O{\textquoteright}Brien, D. Hover, M. E. Schwartz, V. Bolkhovsky, X. Zhang, W. D. Oliver, I. Siddiqi. A near-quantum-limited Josephson travelling-wave parametric amplifier. \textit{Science} \textbf{350}, 307 (2015).

\item[{[S7]}] J. M. Gambetta, F. Motzoi, S. T. Merkel, and F. K. Wilhelm. Analytic control methods for high-fidelity unitary operations in a weakly nonlinear oscillator. \textit{Phys. Rev. A} \textbf{83}, 012308 (2011).

\item[{[S8]}] U. von L\"upke,  F. Beaudoin, L. M. Norris, Y.  Sung, R. Winik, J. Y. Qiu, M. Kjaergaard, D. Kim, J. Yoder, S. Gustavsson, L. Viola, and W. D. Oliver. Two-Qubit spectroscopy of spatiotemporally correlated quantum noise in superconducting qubits. \textit{PRX Quantum} \textbf{1}, 010305 (2020).

\item[{[S9]}] H. P. Breuer and F. Petruccione, \textit{The Theory of Open Quantum Systems} (Oxford University Press, Oxford, 2002).

\item[{[S10]}] In the main text, we moved to the interaction picture with respect to the free Hamiltonian of the bath $H_{\mathrm{B}}$, and introduced the time-dependent noise operator $B^{(j)}(t)\equiv e^{iH_{\mathrm{B}}t/\hbar} B^{(j)} e^{-iH_{\mathrm{B}}t/\hbar}$.

\item[{[S11]}] A. A. Clerk, M. H. Devoret, S. M. Girvin, F. Marquardt, and R. J. Schoelkopf. Introduction to quantum noise, measurement, and amplification. \textit{Reviews of Modern Physics}, \textbf{82} (2010).

{
\item[{[S12]}] F. Yan, S. Gustavsson, J. Bylander, X. Jin, F. Yoshihara, D. G. Cory, Y. Nakamura, T. P. Orlando, and W. D. Oliver. Rotating-frame relaxation as a noise spectrum analyser of a superconducting qubit undergoing driven evolution. \textit{Nature Communications}, \textbf{4} 2337 (2013).

\item[{[S13]}] P. Krantz, M. Kjaergaard, F. Yan, T. P. Orlando, S. Gustavsson, and W. D. Oliver. A quantum engineer's guide to superconducting qubits. \textit{Applied Physics Reviews} \textbf{6}, 9021318 (2019).

\item[{[S14]}] L. Cywinski, R. M. Lutchyn, C. P. Nave, and S. D. Sarma. \textit{Phys. Rev. B.} \textbf{77}, 174509 (2008).

\item[{[S15]}] J. Bylander, S. Gustavsson, F. Yan, F. Yoshihara, K. Harrabi, G. Fitch, D. G. Cory, Y. Nakamura, J. Tsai, W. D. Oliver. \textit{Nature Physics} \textbf{7}, 565-570 (2011).

\item[{[S16]}] G. A. Alvarez and D. Suter. \textit{Phys. Rev. Lett.} \textbf{107}, 230501 (2011).
}
\end{itemize}
\end{document}